\documentclass[twocolumn,twocolappendix]{aastex631}
\usepackage{xspace}
\usepackage{physics}
\usepackage{amsmath}
\usepackage{float}
\newcommand{\mrk}{Mrk~421\xspace}

\newcommand{\ixpe}{{\it IXPE}\xspace}

\newif\ifrev
\revtrue              % <-- Set to \revfalse to turn OFF highlighting

\begin{document}

%\title{X-ray polarization dynamics of a multi-zone jet emission model}
%\title{X-ray polarization dynamics of a multi-zone jet emission model}
% \title{X-ray polarization dynamics of synchrotron emission from a multi-zone blazar jet}
%\title{Dynamics of the X-ray polarization in a multi-zone synchrotron jet}
%\title{X-ray polarization dynamics of a multi-zone synchrotron jet}
%\title{Dynamics of polarized synchrotron X-ray emission from a multi-zone blazar jet}
\title{Polarization Dynamics of X-Ray Synchrotron Emission from a Multi-Zone Blazar Jet}

% % LEADS
%\author[]{VERITAS collaboration}

%\author[0000-0002-1853-863X]{Manel Errando}
\correspondingauthor{B.~de~Jonge, H.~Zhang}
\email{bdejonge@wustl.edu, haocheng.zhang@umbc.edu}
%\affiliation{Physics Department and McDonnell Center for the Space Sciences, Washington University in St. Louis,
%St. Louis, MO 63130, USA}

\author[0009-0008-0415-7263]{B.~de~Jonge}\affiliation{Department of Physics, Washington University in St Louis, St. Louis, MO 63130, USA}
%\author{P.~L.~Rabinowitz}\affiliation{Department of Physics, Washington University, St. Louis, MO 63130, USA}
\author[0000-0001-9826-1759]{H.~Zhang}\affiliation{University of Maryland Baltimore County, Baltimore, MD, 21250, USA}\affiliation{NASA Goddard Space Flight Center, Greenbelt, MD, 20771, USA}
\author[0000-0002-1853-863X]{M.~Errando}\affiliation{Department of Physics, Washington University in St Louis, St. Louis, MO 63130, USA}
\author[0000-0002-5726-5216]{A.~Gokus}\affiliation{Department of Physics, Washington University in St Louis, St. Louis, MO 63130, USA}
\author[0000-0002-5104-5263]{P.~L.~Rabinowitz}\affiliation{Department of Physics, Washington University in St Louis, St. Louis, MO 63130, USA}
% \author{H.~Zhang}\affiliation{University of Maryland Baltimore County, Baltimore, MD 21250, USA; NASA Goddard Space Flight Center, Greenbelt, MD 20771, USA}

%% Note that the \and command from previous versions of AASTeX is now
%% depreciated in this version as it is no longer necessary. AASTeX 
%% automatically takes care of all commas and "and"s between authors names.

%% AASTeX 6.31 has the new \collaboration and \nocollaboration commands to
%% provide the collaboration status of a group of authors. These commands 
%% can be used either before or after the list of corresponding authors. The
%% argument for \collaboration is the collaboration identifier. Authors are
%% encouraged to surround collaboration identifiers with ()s. The 
%% \nocollaboration command takes no argument and exists to indicate that
%% the nearby authors are not part of surrounding collaborations.

%% Mark off the abstract in the ``abstract'' environment. 
\begin{abstract}

%The origin of particle acceleration models in relativistic jets remains an open question. Broad classes of proposed models include reconnection, turbulence (magnetized or otherwise), and shocks. While different models require different physical scenarios (e.g. overall magnetization, magnetic field geometry, etc.), they have been similarly able to explain the emission from blazars. However, with the launch of \ixpe, the magnetic field structure and evolution can be more directly probed so as to distinguish between models of particle acceleration. In this paper we focus on the dynamic X-ray flux and polarization properties of an \ixpe\ observation of Mrk 421 and compare them to a multi-zone model of either magnetized turbulence or magnetic reconnection. Our model leverages the emission from disparate PIC simulations, representing blazar emission as the superposition of many individual simulations. The procedure for combining the simulations has several physically informed parameters which are surveyed and the compatibility of the model with respect to the aforementioned \ixpe\ observation of Mrk 421 is tested. We find that reconnection nearly reproduces the \ixpe\ results for some parameters and magnetized turbulence struggles to reproduce particular observables. We also show tentative results for a mixed model of turbulence and reconnection, motivating the need for proper three dimensional treatments of ththree of 1ES 1959+650 and two of Mrk 421e simulation environments.

The polarization of X-ray synchrotron emission in blazars directly probes the magnetic field geometry and particle acceleration processes in relativistic jets. We use particle-in-cell simulations of magnetic reconnection and magnetized turbulence, coupled to polarization-sensitive radiative transfer code, to interpret \ixpe observations of \mrk during a high flux state recorded in December of 2023. To evaluate the fitness of the two theoretical scenarios, we rely on a quantitative comparison of the statistical properties of simulated and observed X-ray flux and polarization light curves using five evaluation metrics, rather than attempting to fit individual data points. We propose a turbulence-driven multi-zone model where jet emission is represented as the sum of the radiative output of $N$ independent cells, each described by a particle-in-cell simulation. Comparison of ensembles of simulated Stokes-parameter light curves with \ixpe data shows that magnetic reconnection dominated models provide the best match to the observed X-ray flux and polarization dynamics. The optimal configuration corresponds to $N = 15$ emitting cells, which reproduces the observed amplitudes and timescales of the X-ray flux and polarization variations. Magnetized turbulence models underpredict both the flux and polarization variability. Our results indicate that a multi-zone, reconnection-powered emission scenario can describe the X-ray polarization behavior of \mrk and establish a quantitative framework for testing theoretical models against \ixpe observations of other high-synchrotron-peaked blazars.

\end{abstract}

%% Keywords should appear after the \end{abstract} command. 
%% The AAS Journals now uses Unified Astronomy Thesaurus concepts:
%% https://astrothesaurus.org

\section{Introduction} \label{sec:intro}

Blazars are a class of active galactic nuclei whose relativistic jets point very close to our line of sight. They exhibit highly variable non-thermal emission across the electromagnetic spectrum, indicating extreme particle acceleration in very localized regions \citep{2008MNRAS.384L..19B,2019ARA&A..57..467B}. It is often believed that the radiative output from blazars originates from an unresolved region somewhere between sub-parsec to several parsecs from the central black hole, often referred to as the blazar zone \citep{2008Natur.452..966M}. Although the base of the jet is widely considered highly magnetized at the launching site \citep[e.g.,][]{2011MNRAS.418L..79T}, the physical conditions at the blazar zone are still under debate \citep[e.g.,][]{2016ARA&A..54..725M}. If the jet energy is mostly in the form of bulk kinetic energy, then the variable blazar emission is likely attributed to shock and/or shock-induced turbulence that accelerates non-thermal particles \citep{1978ApJ...221L..29B,2015SSRv..191..519S}. Alternatively, if the blazar zone remains considerably magnetized, magnetic reconnection and turbulence are probably the physical driver of particle acceleration and blazar flares \citep{1992A&A...262...26R,2014ApJ...783L..21S,2014PhRvL.113o5005G}. Earlier works trying to distinguish shock, magnetic reconnection, and turbulence mechanisms often considered a simple homogeneous one-zone model to fit the blazar spectral energy distribution and study the underlying non-thermal particle distributions \citep[e.g.,][]{2019ApJ...887..133B}. But numerical plasma simulations have shown that the three mechanisms can accelerate similar non-thermal particle distributions \citep{2014ApJ...783L..21S}. Instead, the three mechanisms involve very different plasma physical evolution, in particular the magnetic field evolution. Therefore, the polarization of the jet's radiative output and its dynamics, a direct measurement of magnetic field morphology and evolution in astrophysical systems, offer a key observable to probe the particle acceleration mechanisms in blazars.

The launch of the {Imaging X-ray Polarimetry Explorer} (\ixpe) has enabled X-ray polarization studies of blazars. Observations generally find a higher degree of X-ray polarization than in the radio and optical bands in high-synchrotron-peaked (HSP) blazars such as Mrk~421 and Mrk~501, supporting an energy-stratified model \citep{Liodakis_2022,Di_Gesu_2022}. Reports on these early results have suggested that energy stratification favors a shock acceleration scenario. 
% \citet{Zhang2024}
% \citet{Bolis2024} have illustrated that the combined effect of radiative cooling and particle transport in an inhomogeneous blazar zone can explain the \ixpe observations, without being exclusively linked to a specific particle acceleration mechanism. 
However, \citet{2024ApJ...967...93Z} has illustrated the combined effect of radiative cooling and particle transport in an inhomogeneous blazar zone, and \citet{Bolis2024} has considered the jet's geometric effects on the multi-wavelength polarization. In both cases, energy stratification is recovered and neither is exclusively linked to a specific particle acceleration mechanism.
\citet{Di_Gesu_2023} reported a coherent $>360^\circ$ rotation of the X-ray polarization angle in Mrk~421, suggesting that angle rotations can originate from a shock propagating down a helical magnetic field in the blazar zone. However, similar polarization angle rotation patterns can be reproduced in a {magnetic reconnection scenario \citep{2015ApJ...804...58Z,2020ApJ...901..149Z}}. Most importantly, follow-up IXPE observations on multiple HSP blazars have shown very rich X-ray flux and polarization behaviors and correlations. Over time scales of months to years, the degree of polarization of HSPs appears to oscillate---Mrk~421 \citep[$\Pi_\mathrm{X}\sim10\%\mbox{--}15\%$,][]{2024AA...681A..12K}, Mrk~501 \citep[$\Pi_\mathrm{X}\sim6\%\mbox{--}19\%$,][]{2024ApJ...974...50C}---yet a clear correlation between the polarization properties and the flux or spectral state of the blazars has not been identified. Blazars 1ES~0229+200 \citep[$\Pi_\mathrm{X}\sim18\%$,][]{2023ApJ...959...61E} and PKS~2155-304 \citep[$\Pi_\mathrm{X}\sim15\%\mbox{--}31\%$,][]{2024AA...689A.119K} have shown higher levels of X-ray polarization than the other HSPs observed so far by \ixpe. In contrast, 1ES~1959+650 exhibited low levels of polarization \citep[$\Pi_\mathrm{X}\lesssim5\%$,][]{2024ApJ...963....5E}, which were lower than optical polarization levels measured at that time. To remain consistent with the energy stratification scenario, the observations of 1ES~1959+650 suggest that X-ray polarization might vary on timescales shorter than those \ixpe can resolve (hours to days), leading to suppression of the measured X-ray polarization.
%
%, such as \textcolor{blue}{please add a few examples here}. 
Given the spread of observational properties observed so far, it is important to study the collective patterns of X-ray flux and polarization in blazars along with a few interesting events.
%Thus it is very important to study the collective patterns of X-ray flux and polarization in blazars in addition to a few interesting events.

%Several previous works have pioneered the study of collective behaviors of blazar radiation and polarization. 
From the theoretical perspective, there have been several attempts to study the blazar radiative output and its polarization properties under different assumptions of particle acceleration and cooling \citep{2012ApJ...744...30K,2013ApJ...774...18Z,2014ApJ...789...66Z,2021Galax...9...27M,2024ApJ...967...93Z}. 
One very successful model is the Turbulent Extreme Multi-Zone (TEMZ) model first described in \citet{Marscher_2014}. This model considers a shock-induced turbulent scenario. It approximates the turbulent blazar zone as multiple independent turbulent cells with different physical parameters and particle distributions that evolve over time, with the general constraint that the overall distribution of cell parameters follows a typical turbulent spectrum. This model is scale-free, and it can directly use typical blazar zone physical parameters to produce simulated light curves for a number of observables such as the flux and polarization in different energy bands that can be compared with observational data.
%general observational patterns. 
Although the resulting radiation and polarization signatures have been shown to match first-principle-integrated simulations \citep{Zhang_2021}, the time evolution of physical parameters such as the cell magnetic field follows random patterns that preserve the overall turbulence spectrum rather than prescriptions informed by the physics of turbulent plasma. %, and the resulting radiation in each cell is not physically solid.
% Evolution is not physical but follows random patterns that preserve the overall turbulence spectrum when integrating over all cells. (e.g. B field evolution is not controlled by the physics of turbulent plasma). 

\begin{figure*}[htpb]
    \centering
    \includegraphics[width=0.95\textwidth]{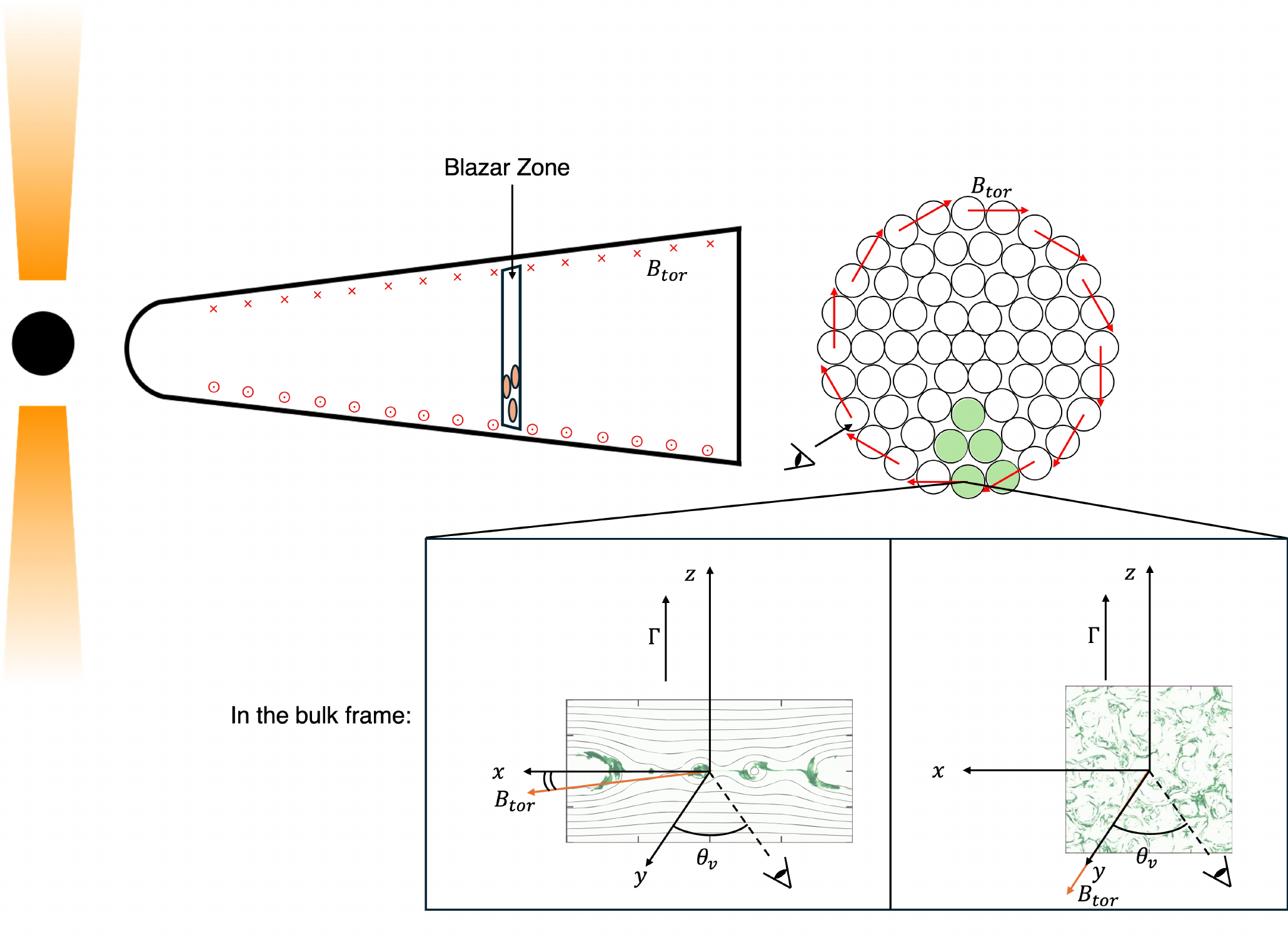}
    \caption{Diagram representing the physical picture of the multi-cell blazar zone model. The jet is discretized into multiple independent cells in an embedded toroidal magnetic field, as seen in the diagram adapted from \citet{Marscher_2014}. 
    %Particle acceleration sites live in a helical field in the jet. The jet is discretized, as seen from the adapted diagram from \cite{Marscher_2014} in the top right of the plot. 
    The model assumes that some subset of the total jet represented by $N$ emitting cells 
%    is responsible for the particle acceleration and emission. 
    responsible for the observed radiative output. Each cell is represented by a particle-in-cell simulation of either magnetic reconnection (left) or magnetized turbulence (right) viewed at some angle $\theta_{\mathrm{v}}$.} %For example, at $\theta_{\mathrm{v}}=0^\circ$ the simulations live in the $x-z$ plane and the observer is along the $y$-axis. 
    %These two pictures are in the frame of the bulk motion of material in the jet, so the observer outside of the jet is oriented near the jet axis as expected for a blazar.
    
    \label{fig:physical_picture}
\end{figure*}

Another approach to study the blazar zone is to use integrated particle-in-cell simulations \citep[PIC, e.g.,][]{2001ApJ...562L..63Z,2008ApJ...682L...5S,2014ApJ...783L..21S} to study the evolution of non-thermal particles coupled to a polarized radiation transfer recipe \citep[][]{1960ratr.book.....C,2014ApJ...789...66Z} to characterize their radiative output. This method has been applied to magnetic reconnection \citep{Zhang_2021} and magnetized turbulence scenarios \citep{Zhang_2023}. The main strength of PIC simulations is their ability to self-consistently track the plasma evolution and particle transport from first principles. Consequently, the resulting radiation and polarization patterns and correlations can be clearly mapped to specific plasma physical processes. In practice, PIC calculations simulate the evolution of particles in small volumes compared to the estimated size of %simulations are known to have very small physical scales compared to 
a realistic blazar zone. Previous studies have shown that PIC results can be consistently extrapolated to large scales \citep{2016MNRAS.462...48S,2024MNRAS.531.4781Z}. %\me{The problem lies in that the initial and boundary conditions of PIC simulations are unlikely to fully capture the inhomogeneous physical conditions on realistic physical scales of the blazar zone. Hence, it is not straightforward to directly compare these first-principle-integrated simulations with observations.}

The X-ray light curves of blazars display flux variability across all measured timescales \citep[e.g.,][]{2021MNRAS.507.5690G,2023MNRAS.526.4040M} with bright flares, intervals of relative quiescence, and periods of time with small-amplitude fluctuations of the flux \citep[e.g.,][]{2022MNRAS.510.4063S}. However, published blazar monitoring campaigns with \ixpe\ have so far not detected significant X-ray flares \citep{Di_Gesu_2022,Liodakis_2022,2023ApJ...953L..28M,Di_Gesu_2023,2023ApJ...959...61E,2024AA...681A..12K,2024ApJ...963....5E,2024AA...689A.119K,2024ApJ...974...50C,2025MNRAS.tmp.1094L}. In this paper, we attempt to characterize these high-flux blazar states without prominent flares that are often observed in HSPs as the radiative output from a multi-cell blazar radiation zone. 
%Here we consider a novel way to marry the strengths of the above two methods. We model the blazar zone similar to the TEMZ model, by many cells with different physical conditions. 
For each individual cell, we assume that the physical conditions become adequately simple so that the radiation and polarization signatures can be extracted from PIC-integrated radiation transfer simulations, ensuring that the radiation and polarization from each cell are physically modeled rather than relying on semi-analytical approximations. As an initial effort, we explore this description of the jet blazar zone to study the dynamics of the X-ray flux and its polarization properties arising from a multi-zone model powered by magnetic reconnection and magnetized turbulence scenarios. 
The simulated emission from multiple cells is subsequently co-added, leading to the X-ray flux expected to be measured by an external observer. 
We compare the simulation results to \ixpe\ data of the blazar Mrk~421 from  December 6 to 22 2023, using a set of metrics that capture the dynamic properties of both simulated and observed X-ray flux and polarization. 
% and compare with \ixpe\ observations. We find that while neither scenarios can fully explain all observed variability patterns, the sum of reconnection simulations performs significantly better than the superposition of magnetized turbulence simulations. In addition, a partially turbulent magnetic reconnection scenario seems to better explain the observations. Section \ref{sec:model} describes the model picture and simulation setup, Section \ref{sec:model_implementation} presents how we implement the model and compare with observations, Section \ref{sec:x} reviews previous attempts to model the same \ixpe\ observation, Section \ref{sec:result} shows the results, and Section \ref{sec:discussion} summarizes and discusses the results.

% Blazars are a class of jetted active galactic nuclei whose relativistic jets are pointed towards the line of sight of the observer. These relativistic jets are magnetized outflows of particles traveling at relativistic speeds. How those particles reach such high energies is the subject of active research. 

% Several acceleration mechanisms have been proposed to explain the implied particle distributions. Broadly, they are shocks, turbulence, and magnetic reconnection. 

\begin{figure*}[htpb]
    \centering
    \includegraphics[width=0.8\textwidth]{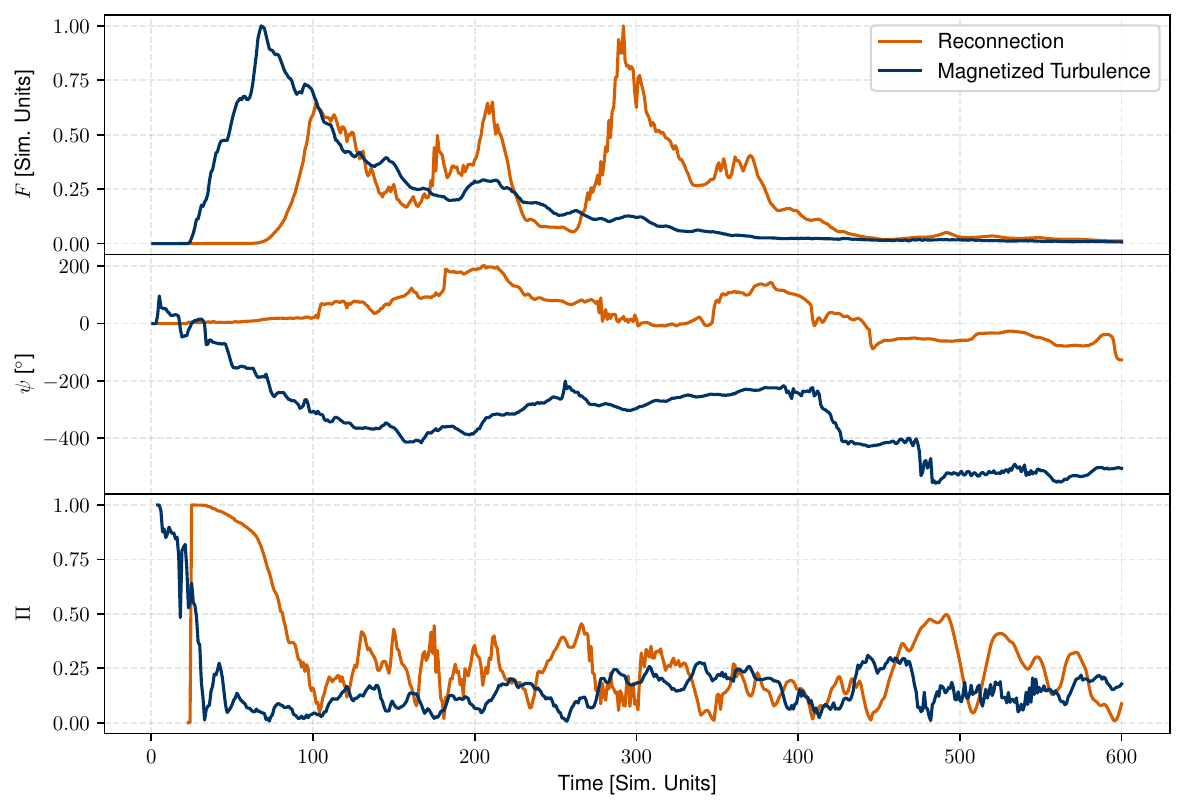}
    \caption{Radiative output for individual magnetized turbulence and magnetic reconnection PIC simulations at $0^\circ$ viewing angles. The plots show the X-ray flux (top), polarization angle (middle), and degree of polarization (bottom).}
        \label{example_sims}
\end{figure*}

\section{Polarization-sensitive PIC-Integrated Radiation Transfer Simulations} \label{sec:sim_details}

A graphical representation of our description of the blazar radiation zone is provided in Figure~\ref{fig:physical_picture}. 
We assume that the magnetic field in the blazar zone consists of a helical field and a turbulent component, as expected in a magnetized jet undergoing magnetic instabilities, as is often seen in global magnetohydrodynamic jet simulations \citep[e.g.][]{Bodo_2020}. The blazar zone moves along the jet direction with bulk Lorentz factor $\Gamma=10$, a typical value in blazars \citep{2009A&A...494..527H}. The line of sight in the observer's frame is chosen to be at $1/\Gamma$ from central axis of the jet. According to relativistic aberration, the line of sight in the comoving frame is then $90^{\circ}$ from the direction of the jet axis. Hence, the bulk Doppler factor is $\delta\equiv\Gamma=10$. We assume that synchrotron emission from the blazar zone can be approximated by the sum of the radiative output from $N$ independent cells. %In this way, $N$ describes how inhomogeneous the blazar zone is. 
Within each cell, we assume that the physical conditions are simple enough 
%so that its variable emission can be extrapolated from a set of 
that we can estimate its radiative output by a set of PIC simulations where the electron-positron plasma is accelerated by either magnetic reconnection or magnetized turbulence. %Since we want to characterize and disentangle the X-ray flux and polarization variability from magnetic reconnection and magnetized turbulence scenarios, we first consider that the $N$ cells are either all Reconnection or all Magnetized Turbulence runs. But we will also show how observable patterns may change if Reconnection and Magnetized Turbulence runs are mixed in the $N$ cells. 
%In light of the different size and free energy budget of cells in the inhomogeneous blazar zone, we further assume that the Stokes parameter $I$, $Q$, $U$ from Reconnection and Magnetized Turbulence runs are weighted by a power-law distribution, $I^{-\alpha}$ \citep{Ciprini_2003}. The different distances from each cell to the plane of sky is implemented in the model as random time delays on top of the Stokes $I$, $Q$, $U$ light curves.
Overall, our setup is a kink-driven magnetized turbulence model in which each cell undergoes magnetic reconnection when the the coherence length of the magnetic field is comparable to the cell size, and turbulence when it is much smaller. Such environments have been validated in 3D magnetized turbulence simulations \citep{2018PhRvL.121y5101C,Comisso2019}. In the following, even though all our scenarios are turbulence-driven, we use ``reconnection'' to indicate simulations with coherence length matching the cell size, and ``magnetized turbulence'' where the coherence length of the magnetic field is much smaller and does not facilitate the formation of current sheets that would lead to reconnection.
%if all cells have coherent lengths comparable to the cell size and undergo reconnection, we call it a reconnection model; if all cells have much smaller coherent lengths, it is called turbulence model; mixture of the two is called hybrid model.}

The radiative output of each individual cell is calculated using magnetic reconnection and magnetized turbulence 2D PIC simulations that  %performed in the $x$-$z$ plane 
use the \texttt{VPIC} code \citep{Bowers2008}. 
%Both the  runs are .
Our PIC simulation setups are described in detail in \citet{Zhang_2021,Zhang_2023}. Nevertheless, we briefly describe their setup. %Here we summarize the key parameters for completeness. 
The simulations assume an electron-ion plasma with a realistic mass ratio $m_i/m_e=1836$. The initial momentum of the particles follows a Maxwell–J\"uttner distribution with uniform density $n_0$ and %temperature $T=T_\mathrm{e}=T_\mathrm{i}$. 
%with uniform density $n_0$ and 
fiducial upstream temperature $T_e=T_i=400 m_e c^2$, where the subscripts indicate electron or ion species. The upstream thermal electron inertial length is then $d_e=d_{e0}\,\sqrt{1+3T_e/(2m_ec^2)}\sim 24.5\,d_{e0}$, where the non-relativistic electron inertial length is $d_{e0}=c/\omega_{pe0}$ and $\omega_{pe0}\equiv\sqrt{4\pi n_0e^2}/m_e$ is the non-relativistic electron plasma frequency. %We choose the electron magnetization factor $\sigma_e=B_0^2/(4\pi n_em_ec^2)=6.4\times 10^5$, corresponding to a total magnetization of $\sigma_0\approx(m_e/m_i)\sigma_e\approx 350$.
The simulation box size is $2L\times L$ and $2L\times 2L$ in the $x$-$z$ plane for reconnection and turbulence, respectively, where $L=48000 d_{e0}\sim 1958d_e$. %The $x$-axis has periodic boundaries for both fields and particles, while the $z$-axis has conductive boundary for fields but reflects particles.
$L$ is resolved by 3072 cells, so that the cell size $\Delta x=\Delta z\sim 0.64d_e$ can resolve the upstream electron inertial length. 
We mimic the synchrotron cooling effect by implementing a radiation reaction force. This term includes two parameters, the cooling strength $C_\mathrm{cool}$ and cooling Lorentz factor $\gamma_\mathrm{cool}$, which are tuned so that the synchrotron cooling break is in the soft X-ray band ($\sim 1~\rm{keV}$). For the magnetic reconnection runs, reconnection starts from a magnetically-dominated force-free current sheet, with anti-parallel magnetic field components with magnitude $B_0$ and a perpendicular guide field with intensity $B_\mathrm{g}=0.2B_0$. To explore a range of initial simulation parameters, we produce a total of 11 reconnection runs exploring two different initial values for $B_\mathrm{g}$, $T$, $C_\mathrm{cool}$, $\gamma_\mathrm{cool}$, and the plasma magnetization parameter $\sigma$.
%$\vect{B}=B_0\tanh(z/\lambda)\hat{x}+B_0\sqrt{\sech^2(z/\lambda)+B_g^2/B_0^2}\hat{y}$, where $B_g=0.2B_0$ is the strength of the guide field, which is the component perpendicular to the anti-parallel components $B_0$. The half-thickness of the current sheet is $\lambda=0.6\sqrt{\sigma_e}d_{e0}$. In addition to this set of fiducial parameter set, we survey two different $B_g$, two $T_e$, two $\sigma_e$, two $C_{cool}$ and two $\gamma_{cool}$: in total 11 magnetic reconnection runs. 
% The magnetized turbulence runs start from a uniform mean magnetic field $B_0\hat{\vect{y}}$ and a spectrum of magnetic fluctuations $\delta\vect{B}$ in the $x$-$z$ plane, with 
The magnetized turbulence runs start from a uniform mean magnetic field $B_0\hat{y}$ and a spectrum of magnetic fluctuations $\delta\vec{B}$ in the $x$-$z$ plane, with 
%, with $\delta B^2_{rms0} \equiv\left<\delta B^2\right>_{t=0}=B_0^2$ and $\delta\vect{B}(\vect{r})=\sum_{\vect{k}}\delta B(\vect{k})\hat{\vect{\xi}}(\vect{k})\exp[i(\vect{k}\cdot\vect{r}+\phi_{\vect{k}})]$, where $\delta B(\vect{k})$ is the amplitude of each wave mode, $\hat{\vect{\xi}}(\vect{k})\equiv i\vect{k}\times\vect{B}_0/|\vect{k}\times\vect{B}_0|$ is the polarization unit vector, and $\phi_{\vect{k}}$ is the wave phase. The wave vector $\vect{k}=(k_x, k_z)$, where $k_x=2\pi m/L_x$ and $k_z=2\pi n/L_z$ for a domain size of $L_x\times L_z$ and $m\in\{-N_x,\cdots-1,1\cdots N_x\}$ and $n\in\{-N_z,\cdots-1,1\cdots N_z\}$. 
$N_\mathrm{x} = N_\mathrm{z}$ %and $N_z$ 
are the number of wave modes along each direction. %We adopt $N_x=N_z$. The wave phases are assumed to be random within 0 and $2\pi$. To ensure that $\delta\vect{B}$ is real, we assume $\delta B(-\vect{k})=\delta B(\vect{k})$ and $\phi_{-\vect{k}}=-\phi_{\vect{k}}$. If each wave mode carries the same power \citep[equal amplitude per mode, similar to][]{Comisso2019}, $\delta B(\vect{k})=\delta B_{rms0}/(2N_x)$. 
We survey two different initial values of $\sigma$, $N_x$, and two initial random seeds that set the amplitude and phase distribution of the turbulence modes, %perturbation profiles 
for a total of 7 magnetized turbulence runs.
% perturbation profiles: two different random seeds that define the amplitude and phase distribution of the turncence modes. 

We post-process the PIC simulations with the \texttt{3DPol} polarized radiation transfer code \citep{Zhang_2014}. The initial magnetic field strength is normalized to $B=0.1~\rm{G}$, which is a typical value for the leptonic scenario \citep[e.g.,][]{Boettcher2013}. 
We bin the particle kinetic energy $(\gamma_e-1)\,m_ec^2$ into 100 steps between $10^{-4}\,m_ec^2$ and $10^6\,m_ec^2$. 
To obtain adequate temporal resolution, we sample the radiative output %We output the above information 
every $\sim 0.0078\,\tau_{lc}$, where $\tau_{lc}$ is the 
light-crossing time in the X direction. 
%to obtain adequate temporal resolution. 
%Under the default resolution, 
%The \texttt{3DPol} code has a 
{The spatial resolution of our simulations is $384\times 192$ radiative transfer cells for magnetic reconnection, and $384 \times 384$ cells for magnetized turbulence.} 
% reconnection 384 x 192
% turb. 384^2
The viewing angle in the comoving frame $\theta_{\mathrm{v}}$ is the angle between the normal vector of the 2D PIC simulation plane and the line of sight; the latter is always perpendicular to the jet direction in the comoving frame (Figure~\ref{fig:physical_picture}). %Due to the 2D nature of our PIC simulations, the physical meaning of viewing from the side is not very clear. Thus the magnetic reconnection and magnetized turbulence runs avoid roughly edge-on ($75^\circ\ \text{and}\ 90^\circ$) viewing angles. 
The \texttt{3DPol} code extracts the Stokes parameters at every time step in each radiative transfer cell, and ray-traces to the plane of the sky to obtain the observed Stokes $I$, $Q$, $U$ parameters that characterize the flux and polarization state of the radiative output of each emitting cell. An example of the temporal evolution of the X-ray flux and polarization contributed by an individual radiating cell is provided in Figure~\ref{example_sims}.

\section{Implementation of a Multi-cell model of the blazar zone} \label{sec:model}

Similar to previous modeling attempts \citep{Kiehlmann_2017, Marscher_2014}, we describe the jet emitting region as $N$ independently-emitting cells of comparable physical size. The emission from each cell is represented by the radiative output of a PIC simulation run powered by either a magnetized turbulence or magnetic reconnection simulation run. Our inhomogeneous blazar zone model is then the sum of the emission from all $N$ cells. 

Each incarnation of our model is represented by the choice of whether the individual cells are powered by magnetic reconnection or magnetized turbulence and three additional parameters: the number of cells $N$ that contribute to the observed emission, the set of viewing angles $\{\theta_\mathrm{v}\}$ that the model can draw from to represent the emission from individual cells, and a power-law index $\alpha$ that determines the distribution of relative weights (or brightnesses) of the emitting cells. In the following, we describe the role that each parameter has in the final model output.

% This has several important parameters that affect the output of this simulation model: 
% \begin{enumerate}[noitemsep,topsep=5pt]
%     \item the number of `cells,' $N$ (i.e. the number of simulations to combine),
%     \item whether Reconnection or Magnetized Turbulence simulations represent cell emission, 
%     \item the index of a power law from which cell flux weights are sampled, $\alpha$, and
%     \item which simulation viewing angles to include in the PIC simulations, $\{\theta_\mathrm{v}\}$.
% \end{enumerate}

% An output of the model is a set Stokes parameter light curves, $I(t),\ Q(t)_,\ \text{and}\ U(t)$. \textit{For clarity, `simulations' will refer to the PIC simulations based on \cite{Zhang_2021, Zhang_2023} whereas the `model' will refer to the superposition of these simulations done in this prescribed way.} As stated more explicitly in section \ref{comparetodata}, we are not fitting a model. The model is a code framework to generate realistic flux and polarization light curves which can then be statistically compared to real polarimetry data from \ixpe. This section will detail the model parameters, how the model is constructed, and how model results are to be compared to observational data.

%\subsection{Methods} \label{methods}
The number of cells $N$ denotes how many cells are significantly contributing to the total observed radiative output, and, in practice, it sets how many PIC simulations runs will be combined together. PIC runs are simulated at viewing angles $\theta_\mathrm{v}$ between $0^{\circ}$ and $60^{\circ}$ in $15^{\circ}$ increments. See Figure~\ref{fig:physical_picture} for the definition of $\theta_\mathrm{v}$. 
The set of viewing angles $\{\theta_{\mathrm{v}}\}$ defines which viewing angles we sample from to construct the output of an incarnation of our model. We only consider sets of adjacent viewing angles. For each viewing angle, there are 11 different magnetic reconnection runs and 6 magnetized turbulence runs that have been simulated and can be sampled from. %are permitted to sample from. $\{\theta_{\mathrm{v}}\}$ will be a subset of the 5 viewing angles ($0^\circ, 15^\circ, 30^\circ, 45^\circ, 60^\circ$) such that each element of the subset is adjacent (e.g. $\{\theta_{\mathrm{v}}\} = \{0^\circ, 15^\circ, 30^\circ\}$ is permitted, but $\{\theta_{\mathrm{v}}\} = \{0^\circ, 30^\circ, 45^\circ, 60^\circ\}$ is not because $0^\circ$ and $30^\circ$ are not continuous in our $15^\circ$ sampling of the viewing angle space).
Once $N$ simulation runs are selected, random time lags are introduced and their individual $I(t), Q(t), \text{and}\ U(t)$ outputs are added together to represent the total radiative output of the model. The flux of each simulation run is normalized by dividing the $I(t)$ light curve by $\int I(t) \mathrm{d}t$ so that each cell initially contributes the same fluence. Additional flux weights, sampled from a power-law distribution with index $-\alpha$, are then assigned to each individual cell to represent variance in individual cell fluence. 
In physical terms, variations in the assigned flux weights represent differences in the available energy of each cell, not differences in their size.
% are sampled from the subset of the simulation database defined by the simulation type and $\{\theta_{\mathrm{v}}\}$, simulations then have random time lags introduced and their fluxes are optionally weighted. The weighting process is defined by the fourth and final parameter, 
We consider $\alpha$ values of 3.5 and 2.5, corresponding to scenarios in which the total radiative output is more or less likely to be dominated by a few bright cells. We also consider the scenario %This value denotes the index of a power law distribution that cell flux weights are sampled from. This, like the sampling of the simulations themselves, is a random process. 
%A part of this fourth dimension of the parameter space represents a weighting scheme 
in which all cells are weighted equally and contribute the same amount of fluence to the final combined light curve.%(i.e. weights are not adjusted by some additional factor sampled from a power law). 
This will be denoted by the shorthand $\alpha=-1$, even though in this case $\alpha$ %. Again, $-1$ 
does not represent a power-law index like positive values of $\alpha$ do. 
%\rev{In the physical picture, differences in cell flux weights correspond to differences in available energy, not cell size.}

%The four dimensional parameter space defined by $N$ (number of cells), the simulation type (whether Reconnection and Magnetized Turbulence populates each cell), $\alpha$ (power law index for flux weights), and $\{\theta_{\mathrm{v}}\}$ (available viewing angles to sample simulations from) is first sliced along the simulation type axis to define two three dimensional parameter spaces corresponding to models whose cells are populated entirely by turbulence simulations and those which are populated entirely by magnetic reconnection simulations. They are compared separately, considering each particle acceleration mechanism's ability to reproduce the statistical properties of Mrk 421 in \ixpe's December 2023 campaign.

For both magnetic reconnection and magnetized turbulence we generate multi-cell model outputs % that time-dependent polarization properties 
by performing a grid search over the following sets of parameters: $N$ = \{5, 15, 25, 35, 45, 55, 65, 75\}, $\alpha$ = \{2.5, 3.5, -1\}, and $\{\theta_{\mathrm{v}}\}$ = $\{\{0^\circ\}, \{15^\circ\}, ..., 
    \{0^\circ, 15^\circ\},$ $\{15^\circ, 30^\circ\}, ..., $\linebreak$ \ \{0^\circ, 15^\circ, 30^\circ, 45^\circ, 60^\circ\}\}$. 

The raw output of the PIC simulation runs includes the temporal evolution of the observed Stokes parameters $I, Q, U$. Stokes $I$ can be considered as a measurement of the observed flux. However, as can be seen in Figure~\ref{example_sims}, the flux output of a simulation run is only provided in arbitrary units, and the time evolution is given in simulation steps. %In order to compare the simulation results to observational data, we need to convert the simulated radiative output into  units of flux as a function of measurable physical time. %A notable obstacle to comparing PIC simulations to observational data is how to extrapolate their results from simulation units to real units. We take the following procedures to compare with IXPE observations. 
To compare the simulation results to observational data, we convert the arbitrary flux units to physical photon fluxes by scaling individual cell radiative outputs so that their sum %For the flux, the cells are simply scaled such that the sum of their radiative outputs that 
represents the total observed flux and has the same average as the observational \ixpe\ flux data. This forces the simulated average flux to match the \ixpe\ average flux, but does not force the flux variability properties to match those of the observational data. The time variability of the simulated flux output will be one of the main features that we will use to discriminate between models. %the variability is left unchanged, which is one of the key features that we compare with observations. 
To convert the simulated time steps into physical time units, we calculate the synchrotron cooling timescale of the electrons responsible for the X-ray radiation as %is estimated with real valued parameters in the simulations to be around $ 
%$\tau_\mathrm{cool, sy} \sim 0.58\,\mathrm{day}\, (\frac{B}{0.1\,\mathrm{G}})^{-2}(\frac{\gamma}{10^5})^{-1}(\frac{\Gamma_\mathrm{bulk}}{10})^{-1}$. 
$\tau_\mathrm{cool, sy} \sim 0.58\,\mathrm{day}\, ({B}/{0.1\,\mathrm{G}})^{-2}\,({\gamma}/{10^5})^{-1}\,({\Gamma_\mathrm{bulk}}/{10})^{-1}$. 
It is expected that this timescale will correspond to the fastest variability timescale that will be seen in the simulations. %frequency in the flux power spectrum where it breaks to flat noise. 
We then construct a power spectrum of the simulated light curve and identify the frequency (in units of inverse simulation steps) % simulation power spectra and the frequency (and corresponding time in simulation time units) 
at which the power breaks to white noise, which corresponds to the fastest variability timescale identified in the simulated light curves. Matching this timescale to the cooling timescale calculated above allows us to estimate the conversion between simulation steps and clock time units in the observer frame.  %This then gives a ratio to extrapolate simulation time units to realistic time units (e.g. days or seconds). 
%This ratio is more specifically defined as $\tau_{sync} \widetilde{\nu}_{break}$ where $\widetilde{\nu}_{break}$ is the median break frequency in the power spectrum fits. 
We find that 600 simulation time steps that we simulate roughly correspond to 80 days. Therefore, in the $\sim$15\,day window of a long \ixpe\ observation can be described with a fraction of a single simulation run.

\section{Observational X-ray polarization data} \label{sec:x}

We test our multi-cell blazar zone model by comparing its predictions to \ixpe\ observations of the blazar Mrk~421, obtained between 2023 December~6 and~22 (MJD~60284--60300) for a total exposure of 512\,ks. The observation spanned 15\,days with three $\sim$36\,h-long gaps.  
%\citep{2024arXiv241019983M}. %Significant temporal variability was detected in both the polarization degree (PD) and polarization angle (PA), with PA fluctuations of $\sim$90$^\circ$ around the jet-axis direction ($\psi \approx 0^\circ$). These variations are consistent with stochastic changes in the magnetic field, possibly arising from turbulence or multi-zone emission. Random-walk simulations \citep{Kiehlmann_2017} were used to model these variations, parameterized by the number of emitting cells ($N_{\rm cells}$) and the number of cells varying per timestep ($n_{\rm var}$). While individual metrics such as the median PD can be reproduced, the simulations have only $\sim$1\% success in matching multiple observed polarization properties simultaneously, indicating that turbulence alone cannot fully explain the data.

\begin{figure*}[tbp]
    \centering
    \includegraphics[width=\textwidth]{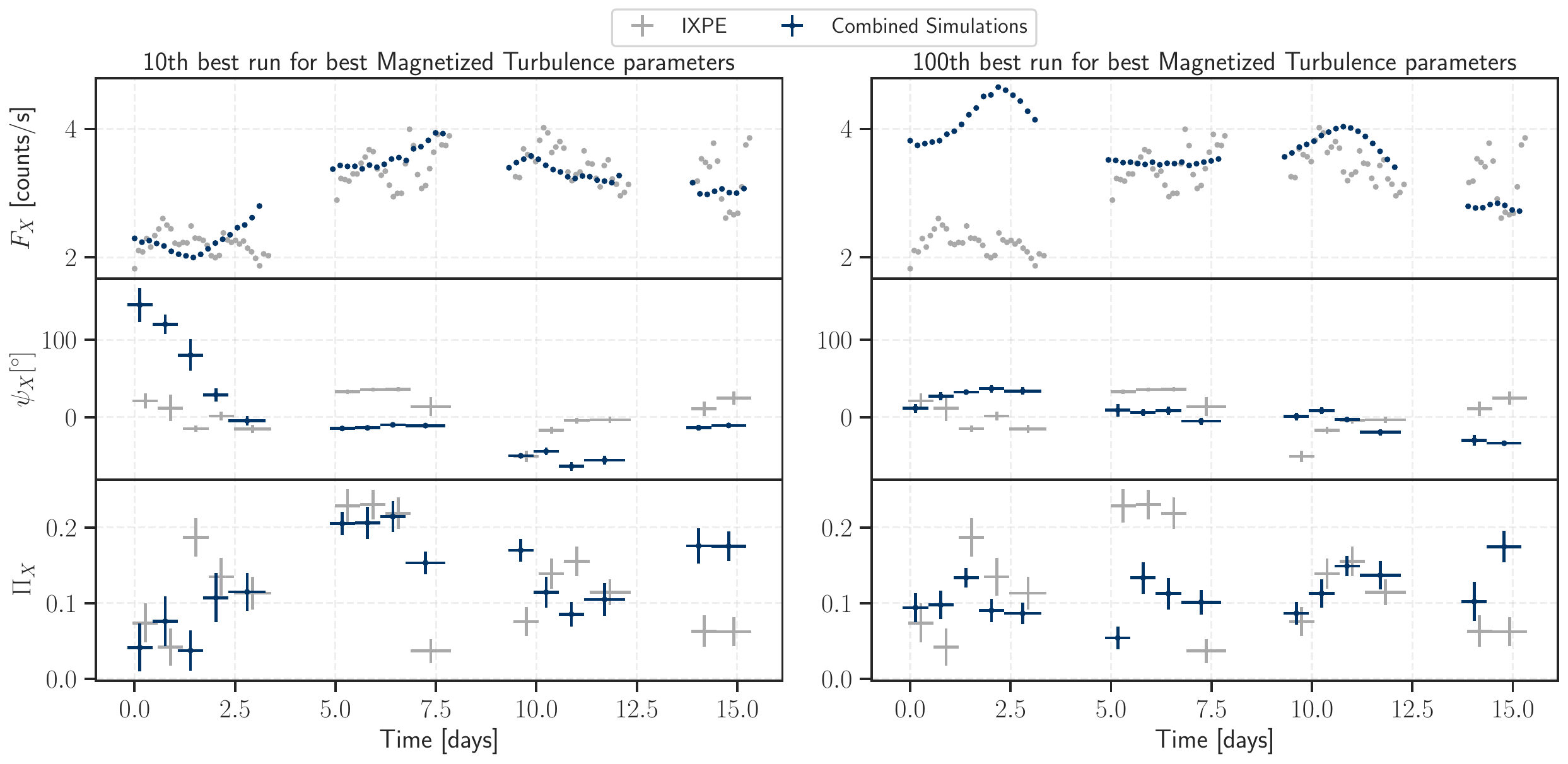}
    \caption{Light curves for the set of turbulence parameters that best approximates all five evaluation metrics simultaneously. This corresponds to parameters $N = 5$, $\alpha = 2.5$, and $\{ \theta_{\mathrm{v}}\} = \{0^\circ\}$. Simulated light curves are compared to the observational \ixpe data. Sorting by the total distance metric, the 10th and 100th best-matching scenarios(out of 1000 simulated outputs) 
    %total outputs which constitute the distribution of outputs) 
    are plotted.}
    \label{bestturblcs}
\end{figure*}

Level~2 \ixpe\ event files, containing event-by-event Stokes parameters ($I$, $Q$, $U$), were obtained from the \texttt{HEASARC} archive and processed with \texttt{ixpeobssim~v30.2.2} \citep{2022SoftX..1901194B,2022ascl.soft10020B}. Source events were selected from a $60''$-radius circular region centered on \mrk, and background from a $150''$--$250''$ annulus, for each detector unit \citep[][]{weisskopf2022}. %Given the background dominance above 6\,keV, analysis was restricted to 2--6\,keV, with Stokes spectra extracted using \texttt{xpbin} and binned to a minimum of 40 counts per bin. Stokes $I$ spectra from all DUs were jointly fit in \texttt{Xspec} with an absorbed log-parabola model, fixing $N_{\rm H}$ to the best-fit value from \swift\ data. The best-fit spectral model was then applied to the $I$, $Q$, and $U$ spectra with an additional \texttt{polconst} component to derive PD and PA. 
The degree of polarization as a function of time is obtained as
\begin{equation}
    {\Pi_{X}}(t) = \frac{\sqrt{Q(t)^2 + U(t)^2}}{I(t)},
\end{equation}
and the polarization angle as
\begin{equation}
    {\psi_X}(t) = \frac{1}{2} \arctan\left( \frac{U(t)}{Q(t)} \right),
\end{equation}
where $I(t)$, $Q(t)$, and $U(t)$ are the background-subtracted Stokes light curves binned over the desired time intervals.

The sensitivity of \ixpe to detect a polarized X-ray flux %an instrument to characterize the polarization state of incoming X-ray photons 
is quantified by the \textit{Minimum Detectable Polarization} (\textit{MDP}), which represents the smallest degree of linear polarization that can be distinguished from statistical noise at a specific confidence level, which is commonly chosen as 99\%. The size of the polarization bins in the December 2023 \textit{IXPE} observation, 15.125 hours in clock time, is motivated by the $MDP_{99}$. For all but one bin, the size of the bins provides sufficient counts such that the $MDP_{99}$ is below the calculated polarization degree, which allows us to maximize the cadence for observing polarization variability. The bins prior to the three telemetry gaps or the end of the observation are simply extended.

\section{Comparison of simulated light curves to X-ray polarization data} \label{comparetodata}

\begin{figure*}[htbp]
    \centering
    \includegraphics[width=\textwidth]{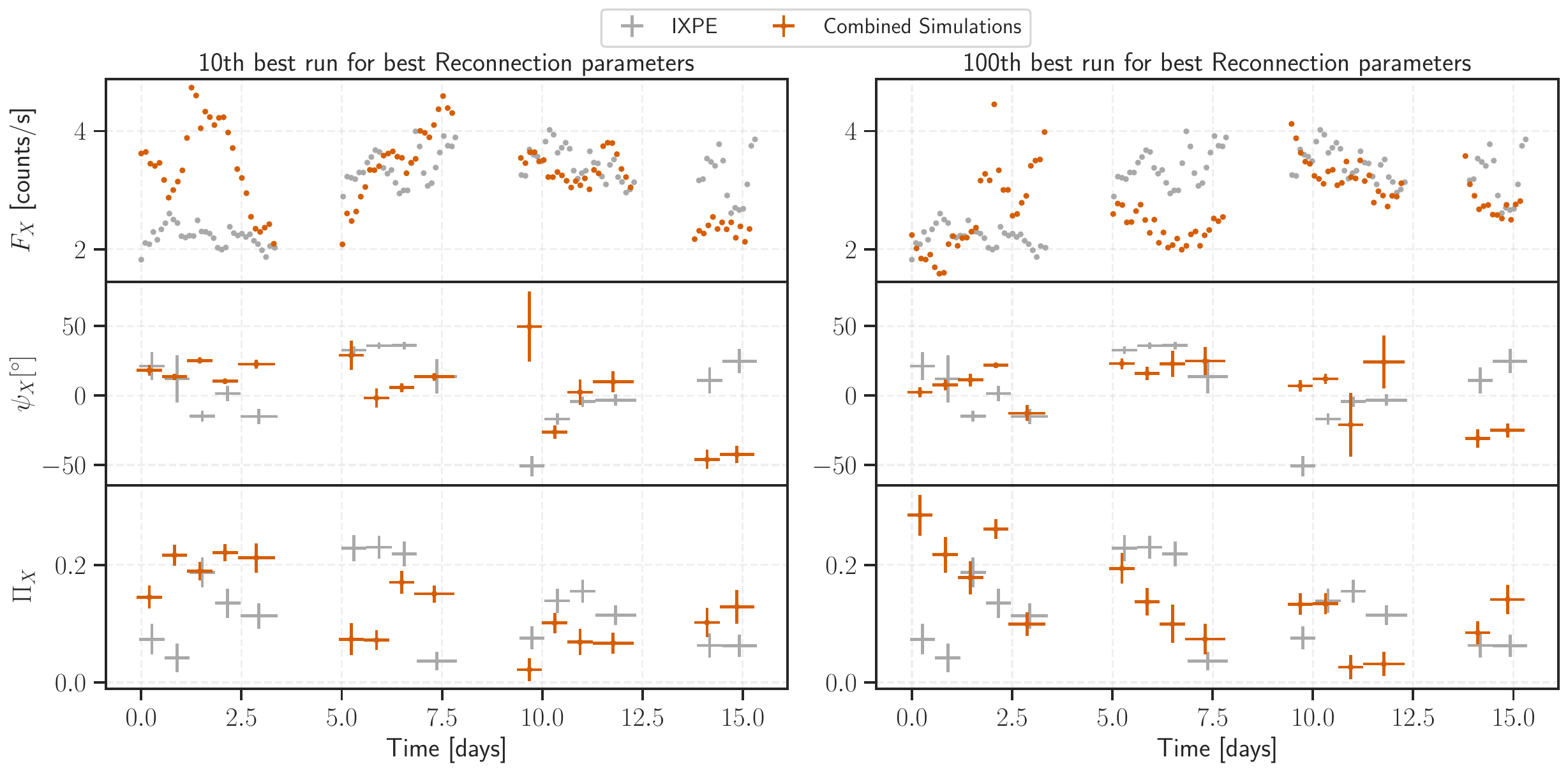}
    \caption{Light curves for the set of turbulence parameters that best approximates all five evaluation metrics simultaneously. This corresponds to parameters $N = 15$, $\alpha = 2.5$, $\{\theta_{\mathrm{v}}\} = \{15^\circ, 30^\circ\}$. Simulated light curves are compared to the observational \ixpe data. Sorting by the total distance metric, the 10th and 100th best-matching scenarios (out of 1000 simulated outputs).}
    %Example output lightcurves for the set of reconnection parameters which best approximates all 5 statistical metrics simultaneously. This corresponds to the parameters $N = 15$, $\alpha = 2.5$, $\{\theta_{\mathrm{v}}\} = \{15^\circ, 30^\circ\}$. Sorting by the defined distance metric, the 10th and 100th best (out of 1000 total outputs which constitute the distribution of outputs) are plotted.}
    \label{bestmr_lcs}
\end{figure*}

Figures \ref{bestturblcs} and \ref{bestmr_lcs} show examples that compare the X-ray flux, polarization angle, and degree of polarization light curves that result from our multi-cell blazar zone simulations with the observational \ixpe data from \mrk described in Section~\ref{sec:x}. 

Given the %inherent random processes in the model and the 
stochastic nature of the plasma dynamics that give rise to the observed X-ray flux and polarization, our aim % and simulated physics, the aim 
is not to exactly reproduce \ixpe\ light curves. Instead, our goal is to find a physical model that produces simulated light curves that reproduce the most salient observable statistical properties. %reproduce a set of statistical metrics constructed from the the flux, polarization angle, and polarization degree light curves. The five metrics considered, largely focusing on dynamic variability, are the
We consider the following evaluation metrics to compare the simulated and observed light curves: variance of the X-ray flux, $\sigma^2(F_X)$;  variance of the polarization angle, $\sigma^2(\psi_X)$; and the mean and variance of the degree of polarization, $\langle \Pi_X \rangle$ and $\sigma^2(\Pi_X)$. Finally, we consider short time scale variance of the X-ray flux by calculating the sum of the variance 
%of the X-ray flux 
evaluated in the time bins used to measure the polarization properties, $\sum_i \sigma^2(F_{X, i})$. This provides an additional metric for quantifying the level of flux variability on short timescales.
%
% \begin{enumerate}[noitemsep, topsep=5pt]
%     \item flux variance, $\sigma^2(F_X)$,
%     \item sum of flux variance in polarization bins, $\sum_i \sigma^2(F_{X, i})$,
%     \item \pa\ variance, $\sigma^2(\psi_X)$,    
%     \item \pd\ mean, $\langle \Pi_X \rangle$,
%     \item \pd\ variance, $\sigma^2(\Pi_X)$,
% \end{enumerate}
All evaluation metrics are weighted by the uncertainty of the measured variables. %inversely weighted by their errors (though flux errors are negligible). The sum of flux variance in polarization bins means that the flux variance is calculated within each polarization bin and then summed.
The average X-ray flux and polarization angle are ignored because the former is artificially matched between observations and simulations, and the latter is a function of an arbitrary geometrical orientation.

\begin{figure*}[htb]
    \centering
    
    \includegraphics[width=\textwidth]{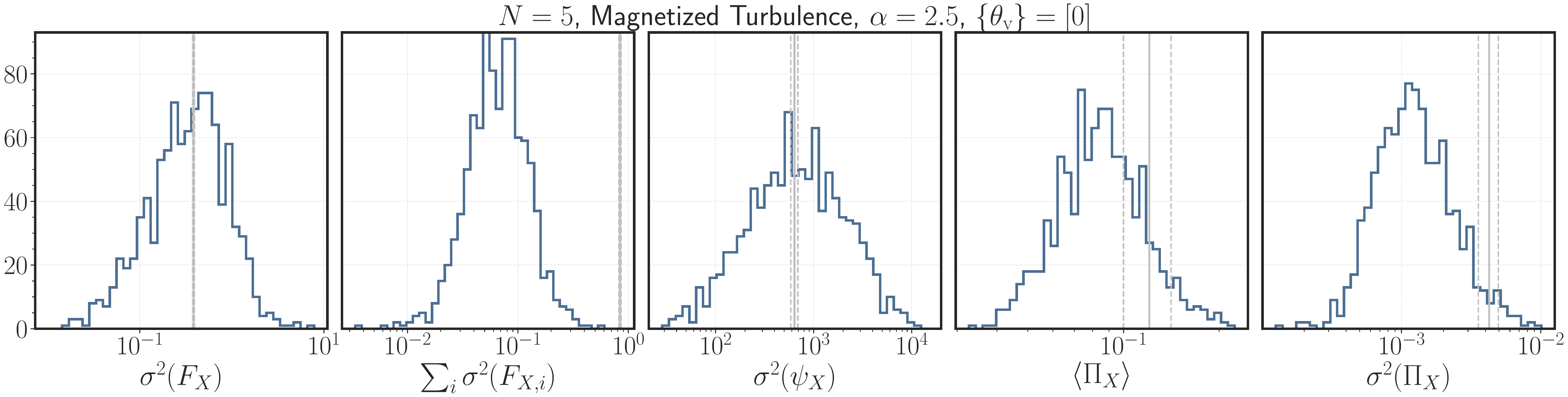}
    
    \caption{Evaluation metric distributions for the set of magnetized turbulence parameters that best approximates the data. The parameters, $N = 5$, $\alpha = 2.5$, and $\{ \theta_{\mathrm{v}}\} = \{0^\circ\}$, represent the combination of 5 magnetized turbulence simulations with flux weights sampled from a powerlaw distribution $\propto A^{-\alpha}$ with $\alpha=2.5$ and only viewed at a viewing angle of $0^\circ$. }%For Magnetized Turbulence, the set of parameters which `best' approximates all metrics (although poorly) also performs best for either flux and polarization subsets of the metrics. The largest deficiency is seen in the ability to replicate short timescale variability.}
    \label{bestturb}
    
\end{figure*}

To evaluate if a 
given set of model parameters can reproduced the statistical properties of the measured \ixpe\ light curves we generate 1000 simulated light curves using the same model parameters. Then, we compare the resulting distribution of evaluation metrics to the observed \ixpe\ values (e.g., Figure~\ref{bestturb}). 
%th eis evaluated by seeing which set of parameters, represented by their distributions over 1000 model outputs, best matches the observed statistical properties. Specifically, with 1000 runs through the discretely sampled parameter space (defined by $N$, $\alpha$, and $\{\theta_{\mathrm{v}}\}$) there are 1000 outputs for each unique combination of parameters. The statistical metrics for each individual result are calculated and form a distribution over 1000 model outputs for each set of parameters. These distributions are then compared to the December 2023 \ixpe\ data of Mrk 421 to see which sets of parameters, if any, can reproduce the observed statistical properties of real data.
% $\frac{m_{ixpe} - m_{sims}}{\sqrt{\sigma^2_{m,sims} + \sigma^2_{m,ixpe}}}$
%
%

\begin{table*}[hbt]
\begin{tabular}{|llll|l|l|l|l|l|}
\hline
\multicolumn{1}{|l|}{$N$} & \multicolumn{1}{l|}{Sim Type} & \multicolumn{1}{l|}{$\alpha$} & $\theta_{v}$ & $\sigma^2(F_X)$ & $\sum_i\sigma^2(F_{X,i})$ & $\sigma^2(\psi_X)$ & $\langle \Pi_X \rangle$ & $\sigma^2(\Pi_X)$ \\ \hline \hline
\multicolumn{4}{|l|}{IXPE}                                                                                                       & 0.384         & 0.840              & 638.2                & 0.1206           & 0.00426              \\ \hline \hline
\multicolumn{1}{|l|}{15}        & \multicolumn{1}{l|}{MR}              & \multicolumn{1}{l|}{2.5}             & 15\_30           & 0.799         & 1.148              & 284.4                & 0.1318           & 0.00308              \\ \hline
\multicolumn{1}{|l|}{5}         & \multicolumn{1}{l|}{MR}              & \multicolumn{1}{l|}{2.5}             & 0\_15            & 1.127         & 0.819              & 1490.7               & 0.1287           & 0.00339              \\ \hline
\multicolumn{1}{|l|}{5}         & \multicolumn{1}{l|}{MR}              & \multicolumn{1}{l|}{2.5}             & 0\_15\_30        & 1.449         & 1.598              & 1057.6               & 0.1454           & 0.00405              \\ \hline
\multicolumn{1}{|l|}{5}         & \multicolumn{1}{l|}{MR}              & \multicolumn{1}{l|}{2.5}             & 0                & 1.153         & 0.668              & 1581.9               & 0.1395           & 0.00400              \\ \hline
\multicolumn{1}{|l|}{5}         & \multicolumn{1}{l|}{MR}              & \multicolumn{1}{l|}{-1}              & 0                & 0.885         & 0.616              & 1764.0               & 0.1243           & 0.00343              \\ \hline \hline
\multicolumn{1}{|l|}{5}         & \multicolumn{1}{l|}{Turb}            & \multicolumn{1}{l|}{2.5}             & 0                & 0.358         & 0.0649              & 665.5                & 0.0843           & 0.00119              \\ \hline
\multicolumn{1}{|l|}{5}         & \multicolumn{1}{l|}{Turb}            & \multicolumn{1}{l|}{2.5}             & 0\_15            & 0.248         & 0.0544              & 119.6                & 0.1306           & 0.00136              \\ \hline
\multicolumn{1}{|l|}{5}         & \multicolumn{1}{l|}{Turb}            & \multicolumn{1}{l|}{3.5}             & 0                & 0.327         & 0.0597              & 735.4                & 0.0787           & 0.00111              \\ \hline
\multicolumn{1}{|l|}{5}         & \multicolumn{1}{l|}{Turb}            & \multicolumn{1}{l|}{3.5}             & 0\_15            & 0.232         & 0.0450              & 117.6                & 0.1263           & 0.00129              \\ \hline
\multicolumn{1}{|l|}{5}         & \multicolumn{1}{l|}{Turb}            & \multicolumn{1}{l|}{-1}              & 0\_15            & 0.222         & 0.0206              & 111.8                & 0.1229           & 0.00122              \\ \hline

\end{tabular}

\caption{Best five sets of model parameters for magnetic reconnection and magnetized turbulence according to the Eq.~\ref{distance} criterion. The first four columns list the parameters of the simulation model. The remaining five columns are the calculated values for the evaluation metrics. For reconnection, it was also required that each metric meet a threshold of 0.5 for the distance in Eq.~\ref{distance}. This was not required for turbulence because it failed to meet any reasonable threshold for all metrics. The metric values calculated for the \protect\mrk\ \protect\ixpe\ data are shown in the first row.}

\label{best_params_table}
\end{table*}

For convenience, we also define the sum of the distances between the observed and simulated values of each metric $m$ normalized by their variance and measured uncertainty:
%To quantify a set of parameter's ability to reproduce the data, the difference between the observed value of a given metric with \ixpe\ and the median value of a particular parameter set's distribution along a metric is normalized by the distribution variance and \ixpe\ error: 
\begin{equation}
    \sum_{m} D_m = \sum_{m}{\dfrac{{|m_{\mathrm{obs}} - m_{\mathrm{sim}}|}}{{\sqrt{ \sigma^2(m_\mathrm{obs}) + \sigma^2(m_\mathrm{sim})}}}} 
    \label{distance}
\end{equation}
where $m$ runs through each of the five evaluation metrics. We use the value of $\sum_{m} D_m$ to evaluate the combined ability of a set of model parameters to describe the observed data across all evaluation metrics. 

\section{Results} \label{sec:result}

\begin{figure*}[hbt]
\centering
    \includegraphics[width=0.45\textwidth]{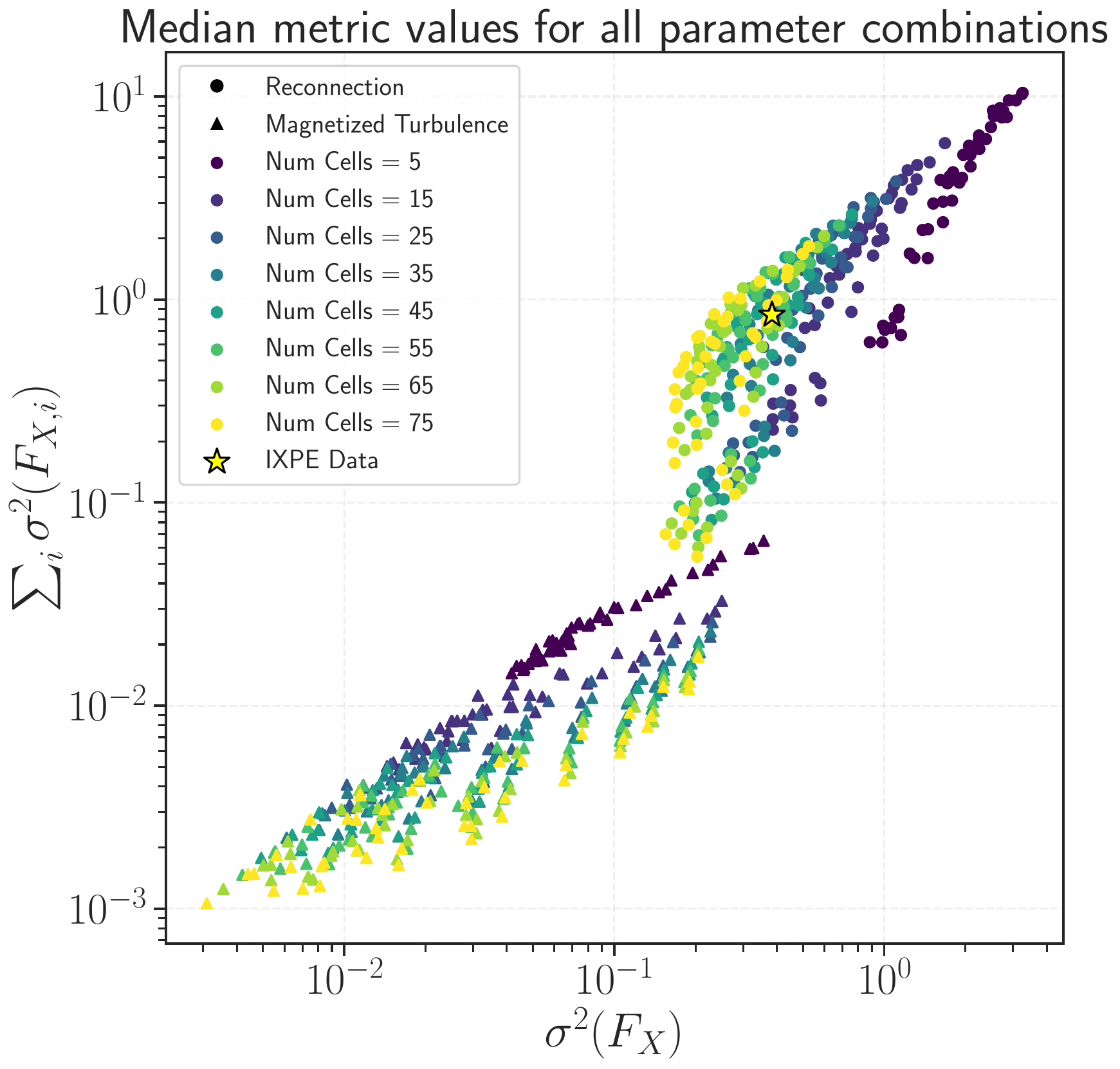}
    \hfill
    \includegraphics[width=0.45\textwidth]{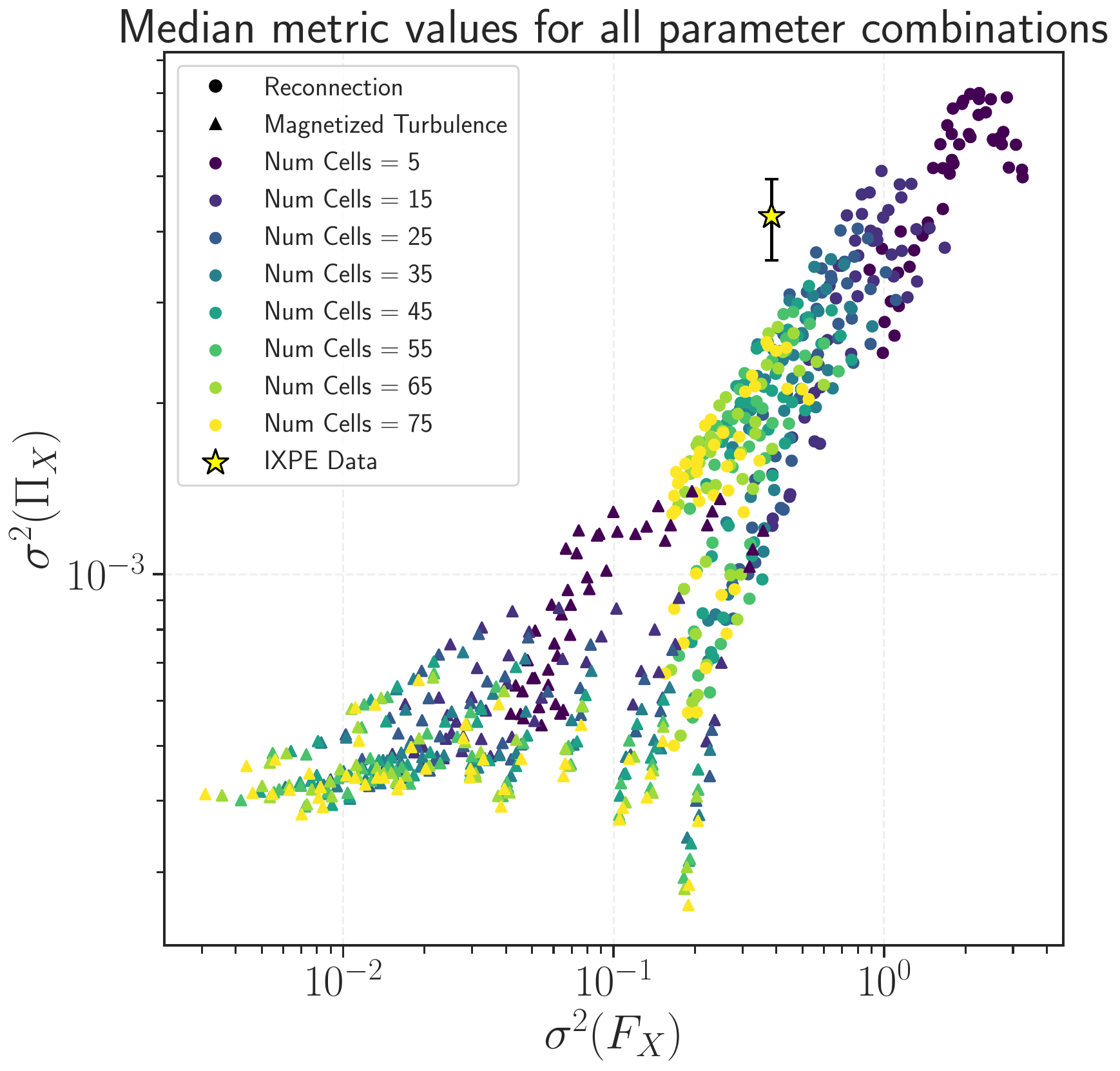}
\caption{Scatter plot showing the  median values of different parameter's distributions for two pairs of metrics: (left) the flux variance against the mean binned flux variance, and (right) the flux variance against the weighted $\Pi_\mathrm{X}$ variance.} %In the former, the projection of the parameter space for the two metrics easily overlaps with the observed \ixpe\ value, whereas in the latter there is no overlap (even with errors on the \ixpe\ value). This shows some tension in matching the variance of both the flux and degree of polarization for a pure 2-D Reconnection simulation.}
\label{fig:2D_metric_plots}
\end{figure*}

First, we evaluate the ability of magnetized turbulence simulations to reproduce the observed X-ray flux and polarization dynamics observed in \mrk. % behavior by Mrk 421 was investigated. By the same distance metric, 
Using the $\sum_m D_m$ criterion described in Eq.~\ref{distance}, the set of turbulence parameters that is closest to describing the observed light curves is $N = 5$, $\alpha = 2.5$, and $\{ \theta_{\mathrm{v}}\} = \{0^\circ\}$. However, as seen in example light curves shown in Figure~\ref{bestturblcs} and the distribution of evaluation metrics shown in Figure~\ref{bestturb}, magnetized turbulence scenarios are not able to reproduce the short-timescale flux variability observed in the \ixpe\ data and captured by the $\sum_i \sigma^2(F_{X, i})$ metric. The magnetized turbulence models also tend to underproduce the overall variability of the degree of polarization (Figure~\ref{bestturb}). 
As illustrated in Figure~\ref{fig:2D_metric_plots}, turbulence models never reach the observed values for several metrics, particularly the short-timescale flux and polarization degree variances, and even their best individual realizations cannot simultaneously reproduce both flux and polarization behavior. While certain simulations can qualitatively match the observed polarization trends, the lack of corresponding flux variability highlights a fundamental limitation of the turbulence scenario in explaining the combined X-ray flux and polarization properties observed in \mrk.

% \begin{figure*}[hbt]
% \centering
% \begin{subfigure}[][][]{0.45\textwidth}
%     \includegraphics[width=\textwidth]{2D_metric_plot_Flux Variance_Binned Flux Variance.pdf}
%     \caption{}
%     \label{fig:flux_var_vs_binned_flux_var}
% \end{subfigure}
% % \hfill
% \begin{subfigure}[][][]{0.45\textwidth}
%     \includegraphics[width=\textwidth]{2D_metric_plot_Flux Variance_Weighted PD Variance.pdf}
%     \caption{}
%     \label{fig:flux_var_vs_PD_var}
% \end{subfigure}

% \caption{Scatter plot showing the  median values of different parameter's distributions for two pairs of metrics: (a) the flux variance against the mean binned flux variance, and (b) the flux variance against the weighted $\Pi_\mathrm{X}$ variance.} %In the former, the projection of the parameter space for the two metrics easily overlaps with the observed \ixpe\ value, whereas in the latter there is no overlap (even with errors on the \ixpe\ value). This shows some tension in matching the variance of both the flux and degree of polarization for a pure 2-D Reconnection simulation.}
% \label{fig:2D_metric_plots}
% \end{figure*}

\begin{figure*}[htbp]
    \centering
    \includegraphics[width=\textwidth]{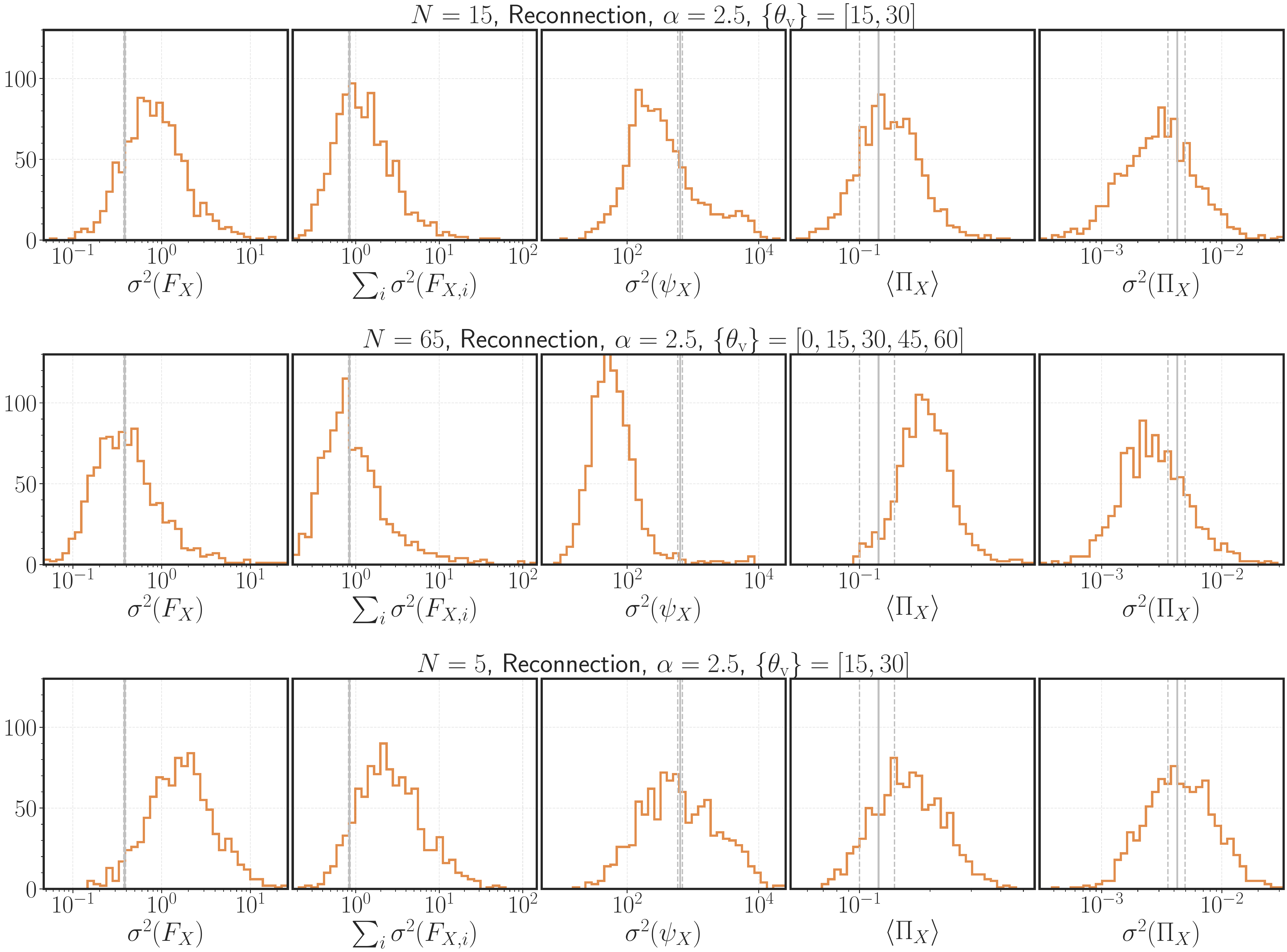}
    \caption{Evaluation metric distributions for the set of magnetic reconnection parameters that best approximates the data. The top panels shown the results for the set of parameters that best combined fit considering all evaluation metrics. The middle panel shows the set of parameters that best match the flux evaluation metrics, while the bottom panel shows the model that best fits the polarization properties of \mrk in December 2023. % figure shows the distribution of model outputs for the set of reconnection parameters which, from top to bottom, best approximates all 5 metrics simultaneously, just the 2 flux metrics, and just the 3 polarization metrics. 
    The metric values for \ixpe\ are shown in dark gray, with their 68\% confidence-level uncertainties plotted in adjacent dashed lines. Note that the histograms are evenly-binned in log space. The set of parameters found to best approximate the observational data is $N = 15$, $\alpha = 2.5$, $\{\theta_{\mathrm{v}}\} = \{15^\circ, 30^\circ\}$. %To better match the two flux variances, more cells (65) and all viewing angles are preferred. Conversely, to better match the three polarization metrics, a small number of cells (5) is preferred.%There is tension is matching the polarization angle variance and flux variance, where matching one variance by (mainly) adjusting the number of cells hurts another variance.
    }
    \label{bestmr}
\end{figure*}

Using the same $\sum_m D_m$ criterion, we find that magnetic reconnection models best reproduce the statistical properties of the observational data %Per the distance $D$ defined in equation \ref{distance}, it was found that the set of parameters corresponding to 
when the total radiative output is described as the sum of 15 cells seen at viewing angles of $15^\circ$ and $30^\circ$ and weighted following a power-law distribution with index $-2.5$ ($N = 15$, $\alpha = 2.5$, $\{\theta_{\mathrm{v}}\} = \{15^\circ, 30^\circ\}$).
%is the most consistent with the \ixpe\ data. 
The evaluation metrics for 100 examples of light curves obtained with this combination of parameters are shown in the top panel of Figure~\ref{bestmr} together with the observed values of the metrics for \mrk\ in December 2023. 
The middle and bottom panels show sets of model parameters that are found to best reproduce the flux or the polarization evaluation metrics, respectively. 
%As illustrated in the middle and bottom panels of Figure~\ref{bestmr}, matching one variability metric—such as the flux variance—does not necessarily reproduce others, like the polarization degree variance. 
There is a clear tension between reproducing the flux and polarization behavior: flux variability is best matched with a large number of emitting cells and higher viewing angles, whereas polarization metrics favor few cells and smaller viewing angles.
The five best sets of parameters are shown in Table~\ref{best_params_table} and show a clear preference for %. These parameter sets were chosen by requiring that $D < 0.5$ for each metric and then sorting by lowest $\sum_mD_m$. The shared characteristics are 
low $N$, $\alpha=2.5$, and low viewing angles. 
Figure~\ref{bestmr_lcs} shows examples of the output of the simulated light curve for the best set of model parameters compared to the observational data. %To visualize what an individual light curve output for the best parameter combination looks like, Figure~\ref{bestmr_lcs} plots two samples from the 1000 outputs which do fairly well (the 10th best and 100th best by total percent error on the metrics).

\begin{figure*}
    \centering
    \includegraphics[width=\textwidth]{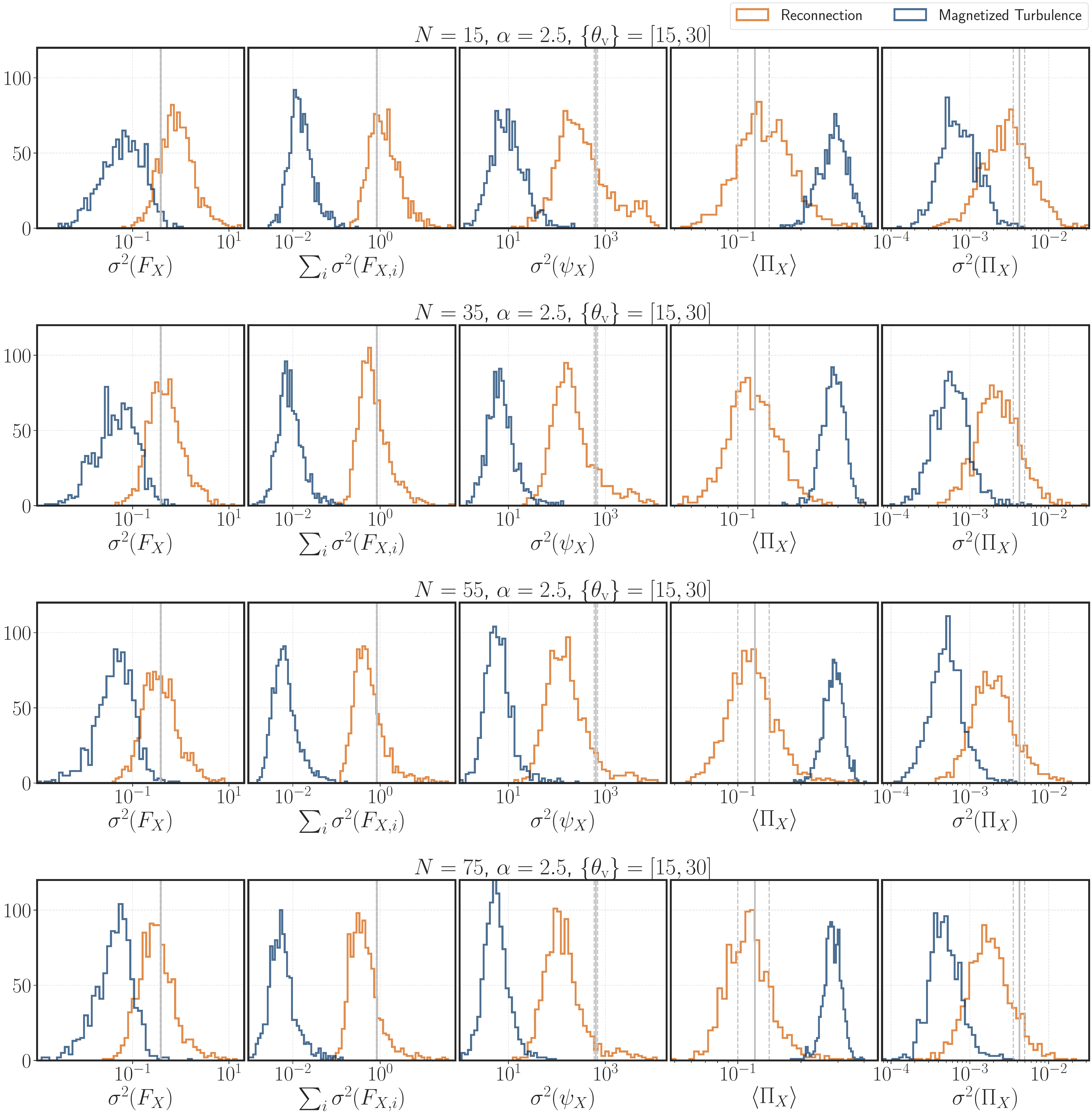}
    \caption{Evaluation metric distributions showing the effect of changing the number of emitting cells in the model. Other parameters are set to their best-fit value for magnetic reconnection simulation runs. Increasing the number of cells reduces the overall variability.}
    \label{fig:n_cells_scan}
\end{figure*}

\begin{figure*}
    \centering
    \includegraphics[width=\textwidth]{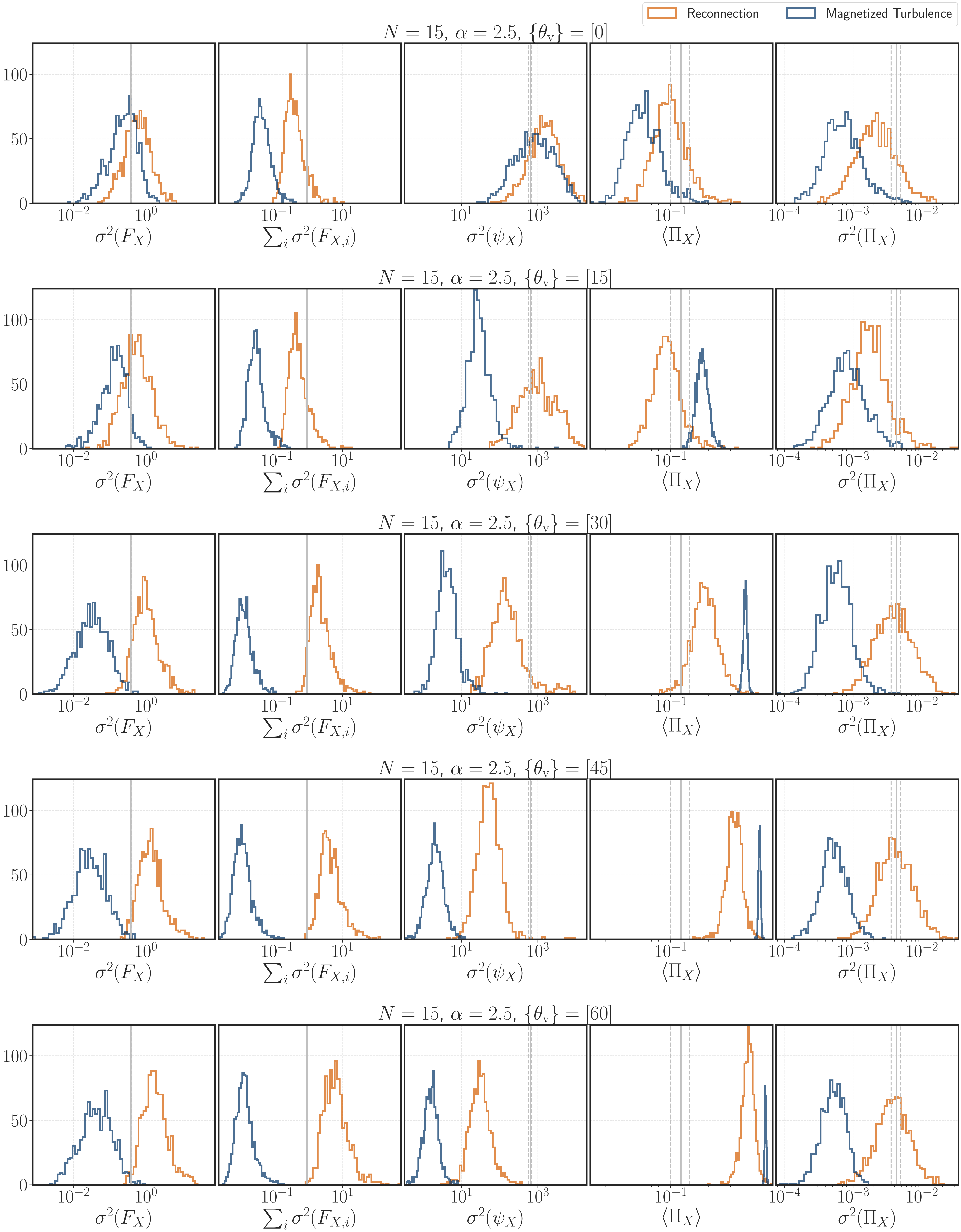}
    \caption{Evaluation metric distributions showing how changing the viewing angle affects the output of the simulations. % are varied to show the affect of the of adjusting the parameter on the distributions produced by the code. 
    Given the large number of possible combinations of $\{ \theta_{\mathrm{v}} \}$, only values with a single viewing angle are considered.} % The effects of the viewing angle are nuanced and different for each metric. They are mostly a function of viewing more or less emission from the more ordered guide field compared to the more turbulent reconnection plane.}
    \label{fig:angle_scan}
\end{figure*}

Given the success of the magnetic reconnection model in producing realistic synthetic light curves, we explore how the flux and polarization dynamics change with changing model parameters. %We restrict the discussion to recone Reconnection (though Magnetized Turbulence results are still plotted). 
First, we consider the effect of changing the number of cells by fixing the other parameters to the best fit values for magnetic reconnection and plotting the distributions of metrics for various values of $N$. The results are shown in Figure~\ref{fig:n_cells_scan}. As expected, when only a few emitting cells contribute, one bright cell can dominate the radiative output and cause large flux fluctuations. As the number of cells increases, their individual fluctuations average out, leading to a suppression of the overall flux variance. For the same reason, the variance of the degree of polarization and polarization angle also decrease with an increasing number of cells. 
%increasing the number of cells reduces the level of flux variability, The effect seems to be that increasing the number of cells reduces variances for the flux, polarization angle, and polarization degree. This makes sense as the effects of outliers become suppressed relative to the growing number of cells.

 %For this reason the effect of changing the power law index is not shown. 

%Similarly, we consider the effect of varying the viewing angle. This is shown in 
Figure~\ref{fig:angle_scan} shows the effect of changing the viewing angle $\theta_\mathrm{v}$ in our physical model while keeping all other parameters the same. We refer the reader to Figure~\ref{fig:physical_picture} for a visualization of the geometrical meaning of the viewing angle in our model setup. %The effect of changing viewing angle is more nuanced than for previous parameters, and to describe it will rely on the picture in Figure~\ref{fig:physical_picture}. 
A viewing angle of $\theta_\mathrm{v} = 0^\circ$ corresponds to a line of sight perpendicular to the magnetic reconnection plane, so the observed emission arises from the combined radiation of all reconnection plasmoids propagating perpendicular to the observer’s line of sight. As $\theta_\mathrm{v}$ increases, the reconnection plane moves closer to the line of sight, and it is more likely that individual plasmoids (with a higher doppler boost toward the observer) dominate the radiative output. As such, when we observe the variance of the X-ray flux in Figure~\ref{fig:angle_scan}, we see that the overall flux variability and the variance at short time scales increase with increasing viewing angle. 
%We observe that the variance of the X-ray flux First we consider both flux variances. As the viewing angle increases, the flux variances increase. 
%We interpret this as progressively seeing plasma fluctuations less uniformly, with higher viewing angles having more emphasized time delays. 
%Next, we consider the polarization angle variance. As the viewing angle increases, the polarization angle variance decreases. 
The polarization properties follow a trend of increasing degree of polarization and decreasing variance of the polarization angle as the viewing angle increases and moves closer to the reconnection plane. The polarization angle is a reflection of the direction of the magnetic field. In the magnetic reconnection simulations, the $x-z$ plane contains more turbulent fields in which reconnection occurs. Conversely, the more stable guide field (the $y$-component of the toroidal field) is along the $y$ axis. At higher viewing angles, the observer sees more synchrotron emission from the guide field which is less randomly oriented. As such, the variance in the polarization angle decreases. Also, as the viewing angle increases the average polarization degree increases. This has a similar explanation to the polarizaton angle variance decreasing. The polarization degree is a reflection of the orderliness of the field lines, so observing emission from the turbulent field lines in the reconnection plane would produce lower polarization degrees than polarized emission from the more stable guide field. 

The polarization properties follow a clear trend: as the viewing angle increases and moves closer to the reconnection plane, the degree of polarization increases while the variance of the polarization angle decreases. The polarization angle follows the orientation of the magnetic field. In our magnetic reconnection simulations, the turbulent magnetic fields that trigger reconnection are primarily confined to the $x$–$z$ plane, whereas the more stable guide field corresponds to the $y$-component of the toroidal field. At larger viewing angles, the synchrotron radiation emitted towards the observer’s line of sight becomes more dominated by this ordered guide field, which increases the order of the observed polarization angle and therefore reduces its variance. The observed increase in degree of polarization arises from the same effect: emission dominated by the turbulent reconnection zones produces lower polarization, whereas emission from the the more coherent guide field yields a higher degree of polarization.

\begin{figure*}[hbt]
    \centering
    \includegraphics[width=\textwidth]{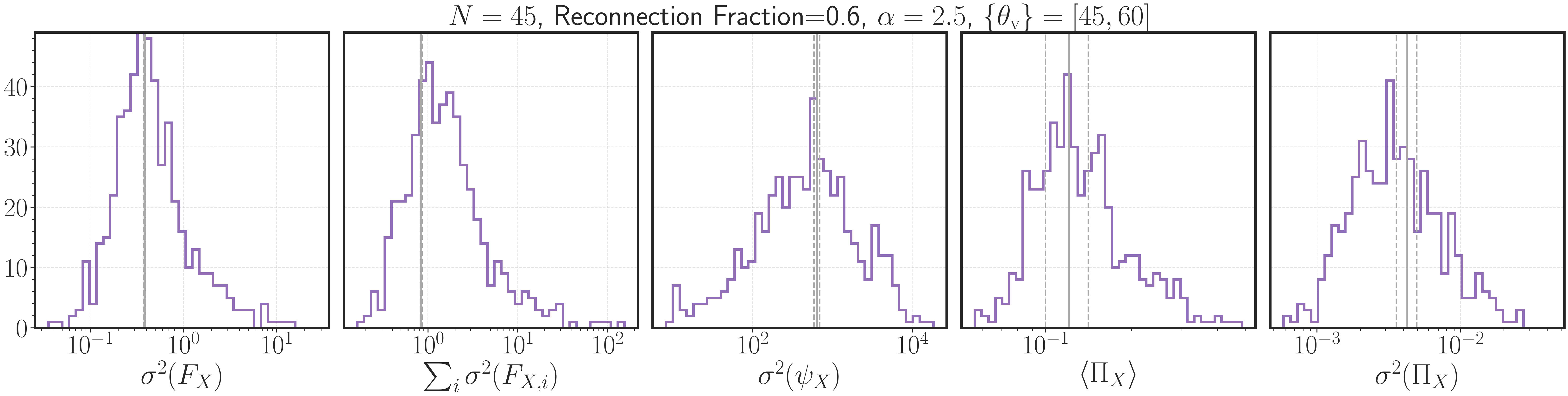}
    \caption{Evaluation metric distributions for 500 model outputs using the combination of magnetized turbulence and magnetic reconnection cells. The model parameters are $N = 45$, $\alpha = 2.5$, $\{\theta_{\mathrm{v}}\} = \{45^\circ, 60^\circ\}$ and a magnetic reconnection Fraction (meaning what fraction of the $N$ simulations are magnetic reconnection, with the remaining simulations being magnetized turbulence) of 0.6 or 60\%} %Their is easy overlap between the distributions and the observed values, though the model may not be completely physical. }
    \label{fig:5_histograms_mixed}
\end{figure*}

\begin{figure}[]
    \centering
    \includegraphics[width=0.95\columnwidth]{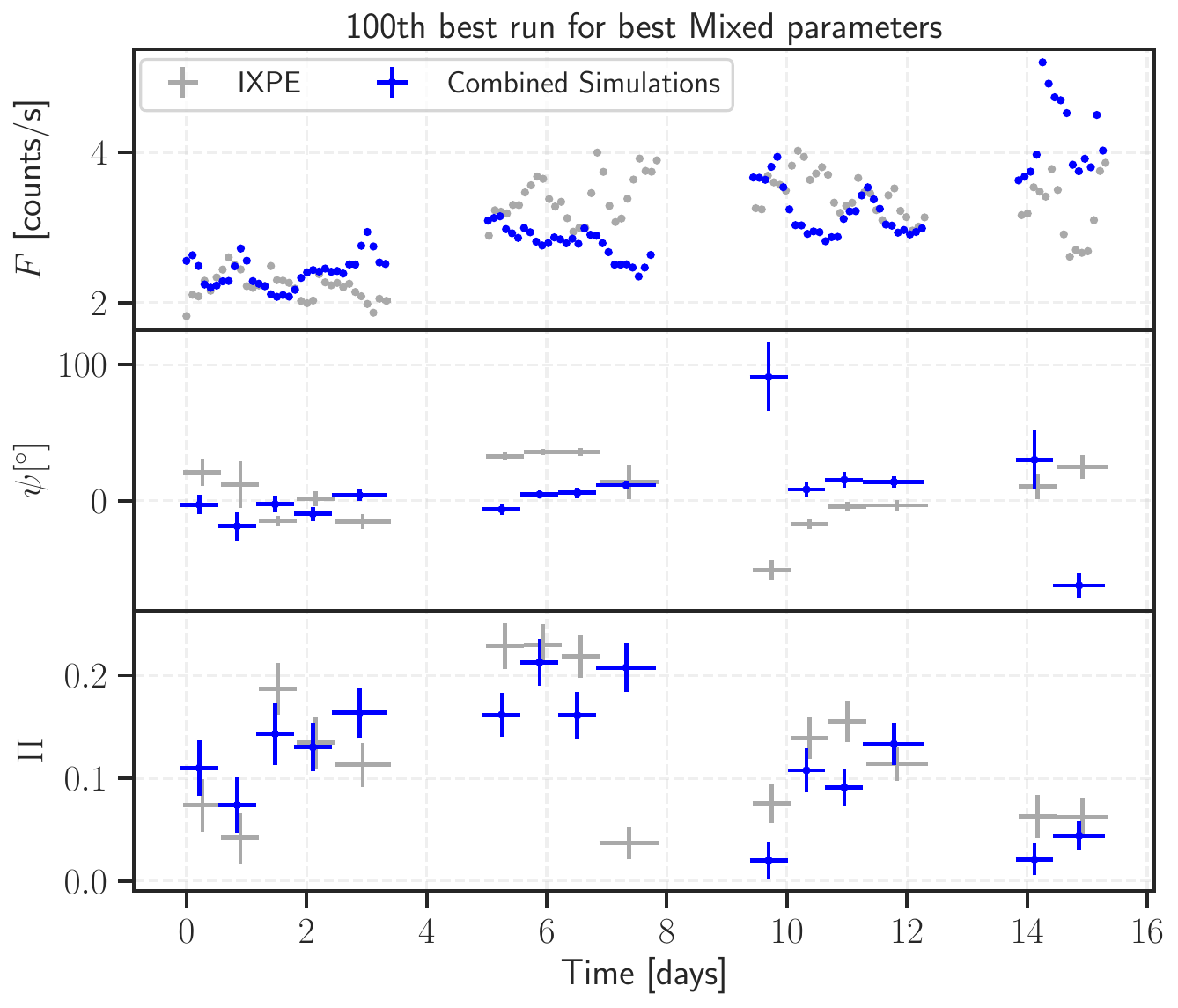}
    \caption{Example output light curve for the mixed magnetic reconnection and magnetized turbulence model. The corresponding parameters are $N = 45$, $\alpha = 2.5$, $\{\theta_{\mathrm{v}}\} = \{45^\circ, 60^\circ\}$. The new parametrization of the reconnection fraction denotes that 60\% of cells are populated by magnetic reconnection simulations and the remaining 40\% are populated by magnetized turbulence.} %This new dimension of the parameter space was sampled at (0, 0.2, 0.4, 0.6, 0.8, 1.0) where 0 represents the pure magnetized turbulence model and 1.0 and pure reconnection.}
    \label{fig:mixed_lcs}
\end{figure}

As shown in Figures~\ref{fig:5_histograms_mixed} and ~\ref{fig:mixed_lcs}, combining magnetic reconnection and magnetized turbulence yields results that closely reproduce the observed flux and polarization behavior. Magnetic reconnection is almost consistent with the data, but there is tension between the flux and polarization variances. Conversely, magnetized turbulence models have a much different flux variability profile while maintaining a roughly comparable polarization profile.
A mixed model, where 60\% of the emitting cells are powered by magnetic reconnection and 40\% by magnetized turbulence, produces light curves and metric distributions that align well with the data. Although this hybrid model is not fully physical, as both processes would likely coexist rather than act independently, it highlights the need for 3D PIC simulations that self-consistently capture the interplay between reconnection and turbulence in the blazar zone \citep{Comisso2019}.

% \begin{figure*}[hbt]
%     \centering
%     \includegraphics[width=\textwidth]{5histograms_45_0.6_2.5_4_5___6_0.pdf}
%     \caption{Distribution of 500 model outputs for the parameter set of mixed reconnection and magnetized turbulence models which was most consistent with the data. The parameters are $N = 45$, $\alpha = 2.5$, $\{\theta_{\mathrm{v}}\} = \{45^\circ, 60^\circ\}$ and a Reconnection Fraction (meaning what fraction of the $N$ simulations are Reconnection, with the remaining simulations being Magnetized Turbulence) of 0.6 or 60\%. Their is easy overlap between the distributions and the observed values, though the model may not be completely physical. }
%     \label{fig:5_histograms_mixed}
% \end{figure*}

\section{Discussion and conclusions}
\label{sec:discussion}
In this paper, we create a framework to compare theoretical simulations of the X-ray flux and polarization dynamics in a multi-zone blazar emitting region with observational data, motivated by the availability of time-resolved polarization properties revealed by \ixpe. 
We develop a model that produces Stokes $I(t), Q(t),\text{and}\ U(t)$ light curves from particle-in-cell simulations powered by two different particle acceleration mechanisms: magnetic reconnection and magnetized turbulence. 
To assess the fitness of different theoretical model scenarios and compare their predictions to observational data, we choose five evaluation metrics that measure the variability of the X-ray flux, polarization angle, and degree of polarization, as well as the average degree of polarization. These metrics can be readily evaluated for simulated light curves as well as for observational data, and allow for a simple comparison of the predicted and observed flux and polarization dynamics from theoretical models and real observations. Rather than directly reproducing the observational data, our framework allows us to identify theoretical models that reproduce the statistical properties of the X-ray flux and polarization light curves. 
We perform a grid search of a multi-cell blazar zone emission model where individual cells are powered by magnetic reconnection or magnetized turbulence, and compare the variability properties of generated light curves to an \ixpe\ observation of \mrk, a high synchrotron peaked blazar, during a high, steady flux state in December 2023.
This roughly 15 day \ixpe observation is one of several \ixpe observations of \mrk, which is just one of several HSPs observed by \ixpe. Other HSPs, such as 1ES 1959+650 and Mrk 501, are characterized by general alignment between the polarization angle and jet direction. HSPs are also consistently characterized by some level of energy stratification, or $\Pi_\mathrm{X} > \Pi_\mathrm{optical}$ \citep[for an overview of alignment/misalignment and energy stratification in HSPs, see][Table 1]{Marscher_2024}. The alignment of the polsarization angle with the projected direction of the jet and the energy stratification are consistent with a shock acceleration model, where particles are accelerated in a shock front with an amplified and compressed magnetic field orthogonal to the jet axis, producing polarized X-ray emission by cooling on a relatively well-ordered magnetic field. Particles then cool and diffuse downstream, occupying a larger volume and sampling a less ordered magnetic fields, resulting in optical emission with lower polarization degree. Initial observations of these phenomena went as far as to argue that they implied acceleration occurred in shocks instead of reconnection \citep[e.g.][]{Liodakis_2022}. Additionally, in June 2022, \mrk underwent a relatively smooth linear rotation of the polarization angle over two three-day observations separated by around a day \citep{Di_Gesu_2023}. The authors interpreted this event as further evidence for the shock model, understanding the rotation as sampling different parts of a helical magnetic field as a shock traveled through it.

Despite an initial expectation that energy stratification would distinguish between shock and reconnection models \citep[e.g.][Table 1]{tavecchio2021probingmagneticfieldsacceleration}, it has been shown that a higher degree of X-ray polarization compared to the optical band can arise in a variety of particle acceleration scenarios and may instead be more sensitive to the geometry of the blazar zone in the jet \citep{2024ApJ...967...93Z, Bolis2024}. In addition, magnetic reconnection can produce radiation with polarization angles aligned with the jet direction in some cases \citep{Tavecchio_2018}. Moreover, \mrk\ has displayed an erratic relationship between its polarization angle and jet direction \citep{Di_Gesu_2022, Di_Gesu_2023, Kim_2024}, leaving additional room for a particle acceleration scenarios such as magnetic reconnection open. % whose relationship with the jet direction is more flexible. 

The December 2023 observation of \mrk\ considered in this paper is characterized by significant variability in both the polarization angle and the degree of polarization. It is the stochastic and highly variable nature of the linear polarization angle and polarization degree that we have aimed to reproduce.
Analyses of this data set have been reported \citep{2024arXiv241019983M,2025A&A...695A.217M}. \citet{2024arXiv241019983M} found significant time variability in the polarization properties throughout the observation. Fluctuations of the polarization angle up to $\sim90^\circ$  around the direction of the jet axis (consistent with $\psi = 0^\circ$) were observed. This implies some random walk of the magnetic field due to turbulence or emission from multiple regions. They test a random walk model with the stochastic variability model from \cite{Kiehlmann_2017}. This random walk model simulates the polarized emission by creating $N_\mathrm{cells}$ cells with a randomly-oriented, ordered magnetic field that each contribute equally to the total radiative output. %and an equal amount of emission. 
Another parameter, $n_\mathrm{var}$ determines how many cells are randomly chosen to be varied at each time step in the simulation. The parameter space defined by these two parameters is simulated. % and the polarization properties of the results are compared to the polarization properties of the \textit{IXPE} data. 
They find that individual properties, such as the median polarization degree, can be matched by the simulation. However, the simulation struggles to simultaneously match several polarization properties with a success rate of $\sim1\%$. This leads to the conclusion that turbulence alone cannot explain the observed polarization properties, although some turbulence must be present to reduce the degree of polarization below the theoretical maximum of $\sim 70-75\%$ for synchrotron radiation \citep{1979rpa..book.....R}. 

% These observations show a range of polarization behaviors, but, up through the December 2023 campaign, are usually relatively steady/quiescent flux states. While other HSPs have shown polarization angles which align with the jet direction, Mrk 421 has shown significant misalignment (c.f. \cite{Kim_2024} Figure~5). For other blazars, alignment has been used as evidence for particle acceleration in shocks (although reconnection could still produce alignment, c.f. \cite{Tavecchio_2018}), but Mrk 421's erratic relationship with its jet direction suggests that magnetic reconnection could be the preferred driver of particle acceleration. The December 2023 observation in particular is characterized by significant variability in both the polarization angle and polarization degree. 

%We used a multi-cell blazar zone modeled with particle distributions and dynamics described by particle-in-cell simulations powered by Magnetized Turbulence and magnetic reconnection to assess the suitability of this scenario to describe the \ixpe\ data set on \mrk\ from December 2023.
We modeled a multi-cell blazar zone using particle-in-cell simulations driven by magnetized turbulence and magnetic reconnection to evaluate how it describes the December 2023 \ixpe\ data on \mrk.
% can be combined to reproduce the observed steady flux and dynamic polarization state observed in December 2023. 
%The model is, at a practical level, the superposition of Stokes $I, Q,\ \text{and}\ U$ lightcurves from many component PIC simulations. The associated physical picture is one of many independent reconnection or turbulence events/cells across the jet. 
%Combining the PIC simulations together can be parametrized along dimensions such as the number of simulations combined together and how the fluxes of component simulations are differently weighted. We conduct a grid search of our parameter space to see if particular sets of parameters can produce results which are statistically similar to the observational data. Notably, we do not try to produce results which perfectly fit the data. This is because, even with the same combination of parameters, each output of our model is unique. There are several steps, whether they be sampling simulations or introducing time lags, which are random. As such, the model is tested by calculating various statistical properties for each model output. 1000 lightcurves are generated for each set of parameters, and the distributions of their statistical properties are compared to the observed values. 
Our study demonstrates the importance of evaluating both the X-ray flux and X-ray polarization model predictions when testing theoretical models. Our magnetized turbulence simulations can reproduce the polarization properties of \mrk\ but fail to capture the observed short-timescale X-ray flux variability. 
In contrast, magnetic reconnection models provide good overall match to the data when emission arises from multiple ($\sim$15) independent cells viewed at angles of $\theta_\mathrm{v}=15^\circ$–$30^\circ$. The model reproduces the observed dynamics of the X-ray flux and polarization, with an increasing viewing angle leading to increased flux variability, higher degrees of polarization, and reduced variance of the polarization angle. These results indicate that a multi-zone turbulence-driven blazar emitting region powered by magnetic reconnection can account for the X-ray flux and polarization variability observed in \mrk in December of 2023. The model evaluation framework presented here can be directly extended to other \ixpe\ observations of bright HSP blazars where the X-ray flux and polarization levels allow time-resolved polarization studies on day-scale or shorter timescales.

\facilities{\ixpe.}

\software{astropy \citep{2013A&A...558A..33A,2018AJ....156..123A,2022ApJ...935..167A},  
\texttt{ixpeobssim} \citep{2022SoftX..1901194B,2022ascl.soft10020B},
HEASoft { \citep{2014ascl.soft08004N}}}

\begin{acknowledgments}

\end{acknowledgments}

% \section{Effects of Adjusting Model Parameters}

% \begin{figure*}
%     \centering
%     \includegraphics[width=\textwidth]{5histograms_n_cells_scan.pdf}
%     \label{fig:n_cells_scan}
%     \caption{Using the other best fit parameters for reconnection, the value of the number of cells is varied to show the affect of the of adjusting the parameter on the distributions produced by the code.}
% \end{figure*}

\clearpage\newpage
\bibliography{bibliography}{}
\bibliographystyle{aasjournal}

%% This command is needed to show the entire author+affiliation list when
%% the collaboration and author truncation commands are used.  It has to
%% go at the end of the manuscript.
%\allauthors

%% Include this line if you are using the \added, \replaced, \deleted
%% commands to see a summary list of all changes at the end of the article.
%\listofchanges

\end{document}